\documentclass[12pt]{article}
\usepackage{graphicx} 
\usepackage{epsfig}
\usepackage{amsmath, amsfonts, amsthm,amssymb}
\usepackage{latexsym}
\usepackage{xcolor}
\usepackage{mathrsfs}
\usepackage{natbib}
\usepackage{hyperref}
\usepackage{appendix}
\usepackage{dsfont}
\usepackage[left=3cm,top=4cm,right=3cm,bottom=3cm]{geometry}
\RequirePackage[caption=false]{subfig}
\usepackage{lscape}
\usepackage{rotating}
\usepackage{ulem}

\usepackage{mathabx}
\usepackage{calligra}

\newcommand{\de}{\mathrm{d}}
\DeclareMathOperator{\tf}{\mathfrak{t}}
\newcommand{\M}{\mathds{M}}
\newcommand{\LL}{\mathds{L}}
\newcommand{\x}{\mathbf{x}}
\newcommand{\e}{\mathds{E}}
\newcommand{\R}{\mathds{R}}
\newcommand{\F}{\mathds{F}}
\newcommand{\1}{\mathds{1}}

\DeclareMathOperator{\card}{card}

\DeclareMathOperator{\prob}{Prob}
\newcommand{\si}{\mathds{S}}

\begin{document}

\title{\LARGE \bf{Local Indicators of Mark Association for Spatial Marked Point Processes} }
\maketitle
\begin{center}
{{\bf Matthias Eckardt$^{1}$} and {\bf Mehdi Moradi$^{2}$}}\\
\noindent $^{\text{1}}$ Chair of Statistics, Humboldt-Universit\"{a}t zu Berlin , Berlin, Germany\\
\noindent $^{\text{2}}$ Department of Mathematics and Mathematical Statistics, Ume\r{a} University, Ume\r{a}, Sweden
\end{center}
\begin{abstract}

The emergence of distinct local mark behaviours is becoming increasingly common in the applications of spatial marked point processes.
This dynamic highlights the limitations of existing global mark correlation functions in accurately identifying the true patterns of mark associations/variations among points as distinct mark behaviours might dominate one another, giving rise to an incomplete understanding of mark associations.
In this paper, we introduce a family of local indicators of mark association (LIMA) functions for spatial marked point processes.
These functions are defined on general state spaces and can include marks that are either real-valued or function-valued.
Unlike global mark correlation functions, which are often distorted by the existence of distinct mark behaviours, LIMA functions reliably identify all types of mark associations and variations among points. Additionally, they accurately determine the interpoint distances where individual points show significant mark associations. 
Through simulation studies, featuring various scenarios, and four real applications in forestry, criminology, and urban mobility, we study spatial marked point processes in $\R^2$ and on linear networks with either real-valued or function-valued marks, demonstrating that LIMA functions significantly outperform the existing global mark correlation functions.

\end{abstract}
{\it Keywords:  Function-valued marks; local mark behaviours; linear networks; mark association; mark correlation functions; mark variogram}

\maketitle

\section{Introduction}

Nowadays, within different applications of spatial point processes, including forestry, environmental sciences, criminology, epidemiology, seismology, etc, it is increasingly becoming common to deal with locations of events that are labelled by some point-specific information called marks. In such cases, the focus shifts from solely the spatial distribution of points to the joint distribution of points and marks, aiming to identify any mark associations among points alongside their spatial distribution. The type of such point-specific information leads to the existence of three general categories for marks depending on the application in question, including integer-valued or categorised marks (e.g., different types of crimes within applications of criminology), real-valued marks (e.g., the diameter at breast height for trees in forestry), and object-valued marks which cover any type of mark that does not fit into the previous categories.  
A common example of object-valued marks includes situations where some functions/curves are attached to the points, e.g., daily cycle distance profiles for bike stations over some time period and population growth of cities over a period of some years \citep{Ghorbani2020, eckardt2024function}. Other examples of object-valued marks include situations where marks are compositions, graphs, and manifolds, among other things. A recent overview of the current state of spatial marked point processes is provided in \cite{Eckardt:Moradi:currrent}.

In cases where events are assigned integer-valued or categorical marks, it is common practice to employ cross/dot-type summary statistics to explore the distribution of points having a specific mark around points having another type of mark. These summary statistics include cross/dot-type $K$-functions, pair correlation functions, empty space distribution functions, nearest neighbour distance distribution functions, and $J$-functions; these are currently available for marked point processes in $\R^2$ as well as those on linear networks \citep{moller2004, Baddeley2014, CronieLieshout2016, Eckardt:Moradi:currrent}. If marks are real-valued, one instead employs summary statistics such as Isham's mark correlation function  \citep{isham1985marked},  
Stoyan's mark correlation function \citep{StoyanStoyan1994}, Beisbart's and Kerscher's mark correlation function \citep{Beisbart:2000}, and  Shimatani's and Schlather's $I$ functions \citep{Shimatani:MoranI, Schlather2004}, which, from different points of view, describe the average association among marks. Furthermore, there exist other summary statistics such as the mark covariance function  \citep{DBLP:journals/eik/Stoyan84}, mark variogram  \citep{cressie93, markvar, Stoyan2000}, and mark differentiation function   \citep{20113358596, HUI2014125}, which all, even though having distinct viewpoints, describe the average variation among marks. These aforementioned mark correlation functions are initially introduced for spatial marked point processes in $\R^2$, but later \cite{Eckardt:Moradi:currrent} proposed their counterparts for spatial marked point processes on linear networks. Note that one could always convert real-valued marks into categorical marks based on some rules and then employ cross/dot-type summary statistics. Turning to object-valued marks, the literature has mostly focused on function-valued marks. 
Second-order summary statistics for spatial marked point processes in $\R^2$ with function-valued marks are proposed by 
\citet{Comas2011, Comas2013}, where they generalised Stoyan's mark correlation function for some applications within forestry.
\cite{Ghorbani2020} presented a profound mathematical formulation for spatial point processes with function-valued marks. Most recently, \cite{Eckardt2023MultiFunctionMarks} and \cite{Eckardt:Moradi:STAT} proposed an extension of general mark correlation functions to multivariate spatial point processes in $\R^2$ and network-constrained spatial marked point processes with function-valued marks, respectively.

These aforementioned summary statistics generally explore either the interactions between points with distinct types of marks, mark associations, or mark variations with respect to a sequence of fixed interpoint distances. What they all have in common is a focus on average behaviours, which, in turn, might often fail to capture all types of behaviours due to the domination between different marks, resulting in erroneous conclusions, especially in situations where marks are space-dependent. 
For instance, consider scenarios where, at a fixed interpoint distance, a set of pairs of points exists, and the associations of some pairs may overshadow those of others, masking the true pattern of mark associations among the points \citep{chaudhuri2023trend}. Therefore, it is of great importance to study the contributions of each point to these average behaviours individually in order to uncover all existing dependence structures among marks, leading to a complete understanding of mark behaviours.
 
For unmarked point processes, studying local dependence structures dates back to when \citet{getis1984interaction} and \citet{getis2010second} proposed some second-order-based statistics similar to $K$-functions for stationary point processes in $\R^2$, focusing on local structures within forestry applications. \cite{StoyanStoyan1994} also discussed similar ideas termed individual functions.
On a different perspective, \citet{LISABundlesCressie, CressieCollinsLISA2} focused on the analogues of Anselin's local indicators of spatial association (LISA) functions \citep{Anselin1995} for spatial point process based on product density functions, namely product density LISA functions. These are also applied by \cite{Mateu:Lorenzo:Porcu:LISA} for feature-clutter classification and \cite{Moraga2011} for disease cluster detection.
\cite{Gonzalez:JABES:LISA:2021} discussed local pair correlation functions for feature-clutter classification. Turning to spatio-temporal point processes in $\R^2\times \R$, \cite{Siino:2018} proposed local indicators of spatial association (LISTA), which were later used by \cite{Adelfio2020} to derive goodness-of-fit diagnostics and \cite{d2023locallyCox} to suggest a local version of minimum contrast estimation for spatio-temporal log-Gaussian Cox processes. 

Turning to locally defined methods for spatial marked point processes in $\R^2$ with integer-valued marks, local versions of (in)homogeneous cross/dot-type $K$- and pair correlation functions are proposed to identify local tendencies towards clustering/inhibition for points with specific marks in relation to points with other types of marks \citep[Chapter 7]{Baddeley2015}. Recently, when marks are functions/curves, \cite{d2024local} proposed local inhomogeneous mark-weighted summary statistics for marked point processes in $\R^2$ which are varieties of $K$-functions.
Moreover, generalisations of Getis-Ord $G$ were considered by \citet{Berglund1999, Flahaut2003} with applications to flow data and traffic accidents.
\cite{lincs:2007, lincs:2010} proposed the so-called local indicators of network-constrained clusters (LINCS) to identify local-scale clustering while \cite{Eckardt2021} discussed the so-called local indicator of network association (LISNA) functions. Finally, considering two distinct spatio-temporal point patterns on a single linear network, \cite{DANGELO2021100534} investigated the local differences between their second-order structure.


As of today, the literature lacks local mark correlation functions for spatial marked point processes with real-valued and/or function-valued marks, which can be great tools for uncovering all types of mark associations/variations among points. 
Aiming to fill this gap, we establish a framework for local mark correlation functions. More specifically, we introduce the family of LIMA functions for spatial marked point process on general state spaces. 
In particular, we consider spatial marked point processes where points are either in $\R^2$ or on linear networks, and marks are either real-valued or function-valued.
All proposed methods are designed to provide useful insights into local variations and dependence structures of marks and, thus, enhance our understanding of inherent structural interrelations for complex spatial marked point processes. By exploring various scenarios in our simulation studies alongside four real-world cases, we demonstrate how our proposed LIMA functions outperform existing global mark correlation functions. Specifically, we focus on the following aspects: the probability of committing a type I error, the power of the test, identifying individual points with significant mark associations, determining interpoint distance ranges where these associations hold, and determining the type of significance in terms of positive or negative mark association. These calculations are done using the \textsf{R} \citep{Rcore} programming languages, and all developed LIMA functions will be accommodated in an  \textsf{R} package.


The paper is organised as follows. Section \ref{sec:pre} provides the necessary background, covering global mark correlation functions for spatial point processes in $\R^2$ and on linear networks, with marks being either real-valued or function-valued.
In Section \ref{sec:locals}, we establish the family of LIMA functions for the same contexts, along with their corresponding non-parametric estimators. Section \ref{sec:simu} presents a simulation study where we assess the performance of our proposed LIMA functions under various scenarios where global mark correlation functions fail to detect the type mark associations correctly.
Section \ref{sec:apps} showcases four real-world applications involving point processes in either $\R^2$ or on linear networks, with marks being real-valued or function-valued, in fields such as forestry, criminology, and urban bike usage.
The paper ends with a discussion in Section \ref{sec:diss}.

\section{Global summary statistics for marked point processes}\label{sec:pre} 
\subsection{Preliminaries} 



Consider $\si$ as an arbitrary complete separable metric space, equipped with a Lebesgue measure $|A| = \int_{A} \de u,\ A \subseteq \mathcal{B}(\si),\ $ which $\mathcal{B}$ stands for Borel sets, and a distance metric $d_{\si}(\cdot,\cdot)$. In addition, let $\M$ be a Polish space associated with $\si$, which we refer to as mark space, equipped with an appropriate reference measure $\nu$ on the Borel $\sigma$-algebra $\mathcal{B}(\M)$. Moreover, the Borel $\sigma$-algebra on the product space $\si \times \M$ is denoted by $\mathcal{B}(\si \times \M)$. Throughout the paper, we focus on a marked point process $X=\{(x_i, m(x_i))\}_{i=1}^N,\ N\geq 0$,  on $\si \times \M$ with points $x_i$ in $\si$ and appertained marks $m(x_i)$ in $\M$. 
Formally, $X$ is considered as a random element of the measurable space $(N_{lf}, \mathcal{N})$ of locally finite point configurations $\x = \{ (x_i, m(x_i) \}_{i=1}^{n}, n\geq 0 $. The associated ground process to $X$, i.e., its unmarked version, is a well-defined point process in $\si$ \citep{Daley2003, Chiu2013}. 
Further, $X$ is considered to be simple meaning that $\prob(\card(X\cap (u,m(u)))\in \lbrace 0,1 \rbrace)\stackrel{a.s.}{=}1$ for all $(u,m(u))\in \si\times \M$. In other terms, there are no multiple points per location. The mark distribution will be denoted by $P_M$ with associated mark mean and mark variance $\mu_M$ and $\sigma^2_M$, respectively.

Within the literature on point processes, there are distinct examples of state space $\si$, including Euclidean space $\si=\R^d,\ d\geq 1$, linear networks $\si=\LL$, and the sphere $\si = \alpha S^{d-1}$ of radius $\alpha$. Although the proposed methods apply to all instances of $\si$, we focus on the first two. The Euclidean space $\si=\R^d,\ d\geq 1$ is equipped with the Euclidean distance $d_{\R^d}(u_1, u_2)= || u_1 - u_2 ||,\ u_1, u_2 \in \R^d$, where $|| \cdot ||$ denotes the Euclidean norm, and with the Lebesgue measure $| \cdot |$. The case of $\si=\LL$ needs particular attention. A linear network $\LL = \cup_{i=1}^k \ell_i$ is considered as a union of $k\geq 1$ line segments 
\[
\ell_i = \left[u_i, v_i\right] = \lbrace tu_i + (1 - t)v_i : 0\leq t \leq 1\rbrace \subseteq \R^d,
\]
with $u_i \neq v_i \in \R^d$ such that for any $i\neq j$, the intersection $\ell_i\cap \ell_j $ is either empty or given by their endpoints. Distances over linear networks are measured based on different metrics, such as shortest-path distance. The class of appropriate metrics for point processes on linear networks is called regular distances, which we denote by $d_{\LL}$. 
Furthermore, integration on $\LL$ is done with respect to arc length and is represented by $\int \de_1 u$ \citep{rakshit2017second, cronie2020inhomogeneous, moradi2024summary}.

\subsection{Marked point processes in Euclidean spaces}\label{sec:marks:points:euclid}

Let $X$ denote a marked point process on $\R^d\times \M$. The expected number of points in $A = B \times C \in \mathcal{B}(\R^d \times \M)$ is 
\[
\Lambda(A)
= 
\e \left[ \card( X \cap A)\right]
=
\int_A \lambda \left( (u, m(u) \right)\de u \ \nu(\de m(u)),
\]
where $\lambda(\cdot)$ is the intensity function of $X$, and governs its spatial distribution. If $\lambda(\cdot)=\lambda$, then $X$ is called a homogeneous point process; otherwise, $X$ is said to be inhomogeneous. 
Moreover, $X$ is stationary if its distribution is invariant under translation, meaning that the distributions of $X=\{(x_i, m(x_i))\}_{i=1}^N,\ N\geq 0$, and $X+s = \{(x_i +s, m(x_i)) \}_{i=1}^N,\ s \in \R^d$, are the same.
Under stationarity assumptions $\Lambda(A)$ reduces to $\Lambda(A) = \Lambda(B \times C)=\lambda |B| P_M(C)$. 

For any non-negative measurable function $h$ on $(\R^d\times\M)^n$, applications of Campbell's formula yield 
\begin{small}
\begin{align*}
& 
\e 
\left[
\sum^{\neq}_{(x_1,m(x_1)), \ldots, (x_n,m(x_n)) \in X}
\hspace{-0.8cm}
h
\Big( 
(x_1,m(x_1)), \ldots, (x_n,m(x_n))
\Big)
\right]
\\
= &
\int_{(\R^d\times\M)^{n}}
h
\Big(
(u_1, m(u_1)),\ldots,(u_n,m(u_n))
\Big)
\\
& \times 
\lambda^{(n)}
\Big(
(u_1, m(u_1)),\ldots,(u_n,m(u_n))
\Big)
\prod_{i=1}^n \de u_i \nu(\de m(u_i)),
\end{align*}
\end{small}
where $\sum^{\neq}$ is a sum over distinct $n$-tuples of points in $X$ and $\lambda^{(n)}$ denotes the $n$-th, $n\geq 1$, order product density function of $X$. Heuristically, $\lambda^{(n)}$ can be interpreted as the probability that $X$ has points $(x_i, m(x_i))$ in distinct infinitesimal regions $\de (x_i, m(x_i)) \in \R^d \times \M$.
However, it is important to note that small/large values of $\lambda^{(n)}$ do not imply any association between the marked points. 
Instead, imposing that the intensity functions of all orders are bounded away from zero, associations between the $n$ points $(x_i, m(x_i)), i=1,\ldots, n$ could be revealed using the correlation function $g^{(n)}_{\R^d \times \M}$ defined by
\begin{align*}
& 
g^{(n)}_{\R^d \times \M}
\Big(
(x_1, m(x_1)),\ldots,(x_n,m(x_n))
\Big)
\\
=
&
\frac{
\lambda^{(n)}
((x_1, m(x_1)),\ldots,(x_n,m(x_n)))
}{
\lambda(x_1, m(x_1)) \cdots \lambda(x_n, m(x_n))
}
\\
=
&
\frac{
\lambda^{(n)}_{\R^d}(x_1,\ldots,x_n)
}{
\lambda_{\R^d}(x_1) \cdots \lambda_{\R^d}(x_n)
}
\frac{
f^{(n)}_{\M} (m_1, \ldots, m_n | x_1, \ldots, x_n)
}{
f^{(1)}_{\M}(m_1|x_1) \cdots f^{(n)}_{\M}(m_n|x_n)
}
\\
=
&
g_{\R^d}^{(n)} (x_1,\ldots,x_n) \gamma_{\M}^{(n)} (m_1, \ldots, m_n | x_1, \ldots, x_n),
\end{align*}
where $g^{(1)}_{\R^d \times \M}(x_1, m(x_1))=1$, $f_{\M}$ is the conditional density function of marks given the spatial locations of points, and $g_{\R^d}^{(n)}$ is the $n$-th order correlation function for the ground process \citep{cronie2024discussion, Eckardt2024Rejoinder}. 
Next, we go through global mark summary statistics for marked point processes when $d=2$ and marks are either real-valued or function-valued. In general, these summary statistics, defined for stationary point processes, display the average association among marks as a function of an interpoint distance $r\geq 0$, aiming at uncovering the average space-dependent distributional behaviours for marks.

\subsubsection{Summary statistics for marked point processes with real-valued marks}\label{sec:global:chars}

Here, one needs to employ a test function $\tf_f: \M \times \M \to \R^+$ for constructing global mark correlation functions in their most general form. For every two points $(x, m(x)), (y, m(y)) \in X$, conditional on having an interpoint distance $r$, i.e., $d_{\R^2}(x, y)=r$, let have 
\begin{align*}
    c_{\tf_f}(r)
    =
    \e 
    \left[
    \tf_f 
    \left(
    m(x), m(y)
    \right)
    \Bigl\vert 
    (x, m(x)), (y, m(y)) \in X
    \right],
\end{align*}
which is a conditional expectation in the Palm sense \citep{Chiu2013}. The global mark correlation functions are generally derived from $c_{\tf_f}(r)$.
Under the assumption of mark independence, i.e., \ when $r \rightarrow \infty$, we have
\begin{eqnarray*}
    c_{\tf_f}(\infty)
    =
    \int_{\R}
    \int_{\R}
    \tf_f(m(x), m(y)) \nu(\de m (x)) \nu(\de m(y)).
\end{eqnarray*}
Normalising $c_{\tf_f}(r)$ via $c_{\tf_f}=c_{\tf_f}(\infty)$ gives rise to the $\tf_f$-correlation functions $\kappa_{\tf_f}(r)$ as
\begin{eqnarray*}
   \kappa_{\tf_f}(r)
   =
   \frac{
   c_{\tf_f}(r)
   }{
   c_{\tf_f}
   },
\end{eqnarray*}
whose precise form and interpretation depends on the specification of the test function under study. Note that the normalising factor $c_{\tf_f}$ varies depending on what form the numerator  $c_{\tf_f}(r)$ has. 
Under the assumption of mark independence, $c_{\tf_f}(r)$ coincides with $c_{\tf_f}$. Whence, $\kappa_{\tf_f}(r)$ equals one, and, therefore, any deviations from unity indicate the existence of associations/variations among marks. 
We add that $c_{\tf_f}(r)$ can be rewritten as 
\begin{align}\label{eq:ctf:2ndorderproducts:ratio}
c_{\tf_f}(r)
=
\frac{
\varrho^{(2)}_{\tf_f}
(r)}{
\varrho^{(2)}
(r)},
\end{align}
where the numerator $\varrho^{(2)}_{\tf_f}$ is the $\tf_f$-second-order product density function of
\begin{align*}
&
\alpha^{(2)}_{\tf_f}(B_1, B_2)
\\
&
=
\e
\sum_{\substack{(x,m(x)),\\ (y, m(y))} \in X}^{\neq}
\tf_f 
(
m(x), m(y)
)
\1_{B_1} \{ x \}
\1_{B_2} \{ y \}
\end{align*}
for any $B_1, B_2 \in \mathcal{B}(\R^2)$, where $\1_{B_1} \{ x \} = \1 \{ x \in  B_1\}$ is an indicator functions, and the denominator $\varrho^{(2)}
(r)$ is the second-order product density function of the ground process. 
Recall that we define the mark correlation functions for stationary point processes, and, thus, the second-order product densities only depend on the spatial distance between $x, y \in \R^d$, i.e., $r$. 
For further details, see \citet{Schlather2001, Illian2008,  Eckardt:Moradi:currrent, Eckardt2024Rejoinder}. 

For simplicity, we rewrite $m(x), m(y)$ as $m_1, m_2$. Prominent cases from the literature which describe the average association between marks include
Isham's mark correlation function $\kappa_{mm}^{\mathrm{Ish}}(r)$ \citep{isham1985marked} with the test function $\tf_f(m_1, m_2)=m_1  m_2-\mu_M^2$, 
Stoyan's mark correlation function $\kappa_{mm}^{\mathrm{Sto}}(r)$  \citep{StoyanStoyan1994} with the test function $\tf_f(m_1, m_2)=m_1 m_2$,
Beisbart's and Kerscher's mark correlation function $\kappa_{mm}^{\mathrm{Bei}}(r)$ with the test function $\tf_f(m_1, m_2)=m_1 + m_2$ \citep{Beisbart:2000},
and the $\mathbf{r}$-mark correlation functions $\kappa_{m\bullet}$ and $\kappa_{\bullet m}$ (corresponding to the conditional mark mean) with the test functions $\tf_f(m_1, m_2)=m_1$ and $\tf_f(m_1, m_2)=m_2$, respectively.
Further mark-association-related summary statistics which can be interpreted as the average pairwise mark autocorrelation are Schlather's and Shimatani's $I$ functions $I_{mm}^{\mathrm{Sch}}(r)$ and $I_{mm}^{\mathrm{Shi}}(r)$ with the test functions $\tf_f(m_1, m_2)=(m_1-\mu_M(\mathbf{r}))(m_2-\mu_M(\mathbf{r}))$ and $\tf_f(m_1, m_2)=(m_1-\mu_M)(m_2-\mu_M)$, respectively \citep{ Schlather2004, Shimatani:MoranI}. Here, $\mu_M(\mathbf{r})$ refers to the average of marks for points which stay exactly $r$ units away from each other. 
Turning our attention to the average variation among marks, there exists the mark variogram $\gamma_{mm}(r)$ \citep{cressie93,markvar, Stoyan2000} with the test function $\tf_f(m_1, m_2)=0.5(m_1-m_2)^2$ and the mark differentiation function $\nabla_{mm}(r)$  \citep{20113358596, HUI2014125} with the test function $\tf_f(m_1, m_2)=1-\min(m_1,m_2)/\max(m_1,m_2)$. 
Note that, due to the distinct forms of these mark correlation functions, they each have their own normalising factors. For instance, in the case of Stoyan's mark correlation function $\kappa_{mm}^{\mathrm{Sto}}(r)$, we have
\begin{align*}
    c_{\tf_f}(\infty)
    &=
    \int_{\R}
    \int_{\R}
    \tf_f(m_1, m_2) \nu(\de m_1) \nu(\de m_2)
    \\
    &=
    \int_{\R}
    \int_{\R}
    m_1 m_2 \nu(\de m_1) \nu(\de m_2)
    \\
    &=
    \mu_M^2,
\end{align*}
while in the case of mark variogram $\gamma_{mm}(r)$, we have
\begin{align*}
    c_{\tf_f}(\infty)
    &=
    \int_{\R}
    \int_{\R}
    \tf_f(m_1, m_2) \nu(\de m_1) \nu(\de m_2)
    \\
    &=
    \int_{\R}
    \int_{\R}
    0.5(m_1-m_2)^2 
    \nu(\de m_1) \nu(\de m_2)
    \\
    &=
    0.5
    \Bigg[
    \int_{\R}
    m_1^2 \nu(\de m_1)
    +
    \int_{\R}
    m_2^2 \nu(\de m_2)
    \\
    &- 
    \int_{\R}
    \int_{\R}
    2 m_1m_2 
    \nu(\de m_1)
    \nu(\de m_2)
    \Bigg]
    \\
    &=
    \sigma_M^2.
\end{align*}
It is important to mention that, for Schlather's and Shimantani's $I$ functions, it is standard practice to normalize the test function by $\sigma_M^2$ to create a close analogy to Moran's index $I$ \citep{moran, Shimatani:MoranI}.  
A comparison of the outcomes from different mark correlation functions across various scenarios is presented in \cite{Eckardt2024Rejoinder}. Since these functions study mark associations and variations from different perspectives, their findings will naturally differ.

\subsubsection{Summary statistics for marked point processes with function-valued marks}\label{sec:global:chars:functional}

Setting $\M = \F(\mathcal{T})$ with $\mathcal{T}=(a,b), -\infty\leq a\leq b\leq \infty$, and replacing the marks $m(x_i)$ by the function-valued quantities $f(x_i)(t): \mathcal{T}\subseteq \R \mapsto \R$, leads to spatial point processes with function-valued marks within which $X = \lbrace x_i, f(x_i)(t)\rbrace^N_{i=1},\ N \geq 0$. Following \cite{Eckardt2023MultiFunctionMarks} and writing a generalised test function as $\tf_f:\F(\mathcal{T})\times \F(\mathcal{T})\to \R^+$, one can adapt the mark correlation functions in Section \ref{sec:global:chars} to the present setting and define pointwise mark summary statistics. Setting $f_1=f(x)(t)$ and $f_2=f(y)(t)$, for points $(x, f_1),(y, f_2) \in X$ with an interpoint spatial distance $d_{\R^2}(x,y)=r$, we have
\begin{align}\label{eq:ctffun}
    c_{\tf_f}(r,t)
    = 
    \e 
    \left[
    \tf_f 
    \left(
    f_1, f_2
    \right)
    \Bigl\vert 
    (x, f_1), (y,  f_2) \in X
    \right],
\end{align}
and
\begin{eqnarray}\label{eq:ctffunind}
    c_{\tf_f}(t)
    =
    \int_{\F(\mathcal{T})}
    \int_{\F(\mathcal{T})}
    \tf_f(f_1,f_2)
    \nu (\de f_1)
    \nu (\de f_2).
\end{eqnarray}

Similar to Section \ref{sec:global:chars}, different normalised and unnormalised pointwise mark correlation functions can be defined for the present setting. By taking the conditional expectations $c_{\tf_f}(r,t)$ and $c_{\tf_f}(t)=c_{\tf_f}(\infty, t)$, we can have pointwise $\tf_f$-correlation functions
\[
\kappa_{\tf_f}(r,t)=\frac{c_{\tf_f}(r,t)}{c_{\tf_f}(t)}.
\]
Note that, in this context, pointwise refers to 
the functional argument 
$t$ being fixed. 
Given the pointwise mark characteristics $c_{\tf_f}(r,t)$ and $\kappa_{\tf_f}(r,t)$, the mark characteristics $c_{\tf_f}(r)$ and $\kappa_{\tf_f}(r)$ can then be obtained through integration over $\mathcal{T}$ as 
\[
 c_{\tf_f}(r)
 =  
 \int_\mathcal{T} c_{\tf_f}(r,t)\de t,
\]
and 
\[
\kappa_{\tf_f}(r)
=
\int_\mathcal{T} \kappa_{\tf_f}(r,t)\de t.
\]
As showcases, we can see that,  for all pairs of points $(x, f_1),(y, f_2) \in X$ for which  $d_{\R^2}(x,y)=r$, the unnormalised mark variogram $\gamma_{ff}(r)$ and Stoyan's mark correlation function $\kappa_{ff}^{\mathrm{Sto}}(r)$ are of the forms
\[
\gamma_{ff}(r) =  \int_\mathcal{T}
\e 
\left[
0.5(f_1-f_2)^2
\Bigl\vert 
(x, f_1), (y,  f_2) \in X
\right]
\de t,
\]
and 
\[
\kappa^{\mathrm{Sto}}_{ff}(r)
=
\int_\mathcal{T}
\e 
\left[
f_1 f_2
\Bigl\vert 
(x, f_1), (y,  f_2) \in X
\right]
\de t.
\]

\subsection{Marked point processes on linear  networks}\label{sec:marks:points:nets}


Now, we consider situations where the spatial locations of points are forced to be on a network structure. More specifically, we let $\si$ be a linear network $\LL$ and focus on a marked point process $X_{\LL}$ on $\LL \times \M$. Similar to the case where spatial locations can occur anywhere in space, and following Campbell's formula, for any non-negative measurable function $h$ on $(\LL \times \M)^n$, it holds that 
\begin{small}
\begin{align*} \label{e:CampbellNets}
&
\e
\left[
\sum^{\neq}_{(x_1,m(x_1)),\ldots,(x_m,m(x_m)) \in X_{\LL}}
\hspace{-0.9cm}
h
\Big(
(x_1,m(x_1)),\ldots,(x_m,m(x_n))
\Big)
\right]
\\
=  & \int_{ (\LL \times \M)^n }
h
\Big(
(u_1, m(u_1)),\ldots,(u_n,m(u_n))
\Big)
\\
\times & 
\lambda^{(n)}_{\LL}
\Big( 
(u_1, m(u_1)), \ldots, (u_n,m(u_n)) 
\Big)
\prod_{i=1}^n \de_1 u_i \nu(\de m(u_i)),
\end{align*}
\end{small}
where $\lambda^{(n)}_{\LL}$ is the $n$-th, $n\geq 1$, order product density function of $X_{\LL}$.
Note that, considering the first order of the above equation and letting $h(x, m(x)) = \1 \{ (x, m(x)) \in A \}, A \subseteq \LL \times \M$, we have
\begin{small}
\begin{align*}
&
\e 
\Bigg[
\card( X_{\LL} \cap A) 
\Bigg] 
=
\e 
\left[
\sum^{\neq}_{(x,m(x)) \in X_{\LL}}
h
\left(
(x,m(x))
\right)
\right]
\\
=  & \int_{A}
\lambda^{(1)}_{\LL} (u, m(u)) \ 
 \de_1 u \ \nu(\de m(u)),
\end{align*}
\end{small}
where $\lambda_{\LL}^{(1)}(u, m(u))=\lambda_{\LL}(u, m(u))$ denotes the intensity function of $X_{\LL}$, governing its spatial  distribution over the product space $\LL \times \M$. Currently, there is no appropriate transformation for network structures that can shift points while ensuring they remain on the network after being transformed. Hence, stationary point processes on linear networks cannot be defined in the same manner as those on planar spaces. Imposing that the intensity function $\lambda_{\LL}(u, m(u))$ is bounded away from zero, we have the $n$-th order correlation function as
\begin{align*}
& 
g^{(n)}_{\LL \times \M}
\Big(
(x_1, m(x_1)),\ldots,(x_n,m(x_n))
\Big)
\\
=
&
g_{\LL}^{(n)} (x_1,\ldots,x_n) \gamma_{\M}^{(n)} (m_1, \ldots, m_n | x_1, \ldots, x_n).
\end{align*}
Following \citet{cronie2020inhomogeneous, Eckardt2024Rejoinder, cronie2024discussion}, a point process $X_{\LL}$ is intensity
reweighted moment pseudo-stationary (IRMPS), if 
\begin{align*}
&
g^{(n)}_{\LL \times \M}
\Big(
(x_1, m(x_1)),\ldots,(x_n,m(x_n))
\Big)
\\
=
&
\bar{g}^{(n)}
\Big(
d_{\LL}(u, x_1), \ldots, d_{\LL}(u, x_n), m(x_1), \ldots, m(x_n)
\Big),
\end{align*}
for any fixed $u \in \LL$ and some function $\bar{g}^{(n)}: [0, \infty)^n \times \M^n \to [0, \infty)$. In particular, $X_{\LL}$ is $k$-th order IRMPS if the above equality holds for any $k \geq 2$. If $X_{\LL}$ is homogeneous and $k$-th order IRMPS, it is then $k$-th order pseudo-stationary. Finally, if this holds for any $k \geq 1$, in a way that moments completely and uniquely characterises the distribution of $X_{\LL}$, we call $X_{\LL}$ pseudo-stationary. 

\subsubsection{
Summary statistics for marked point processes with real-valued marks}\label{sec:global:chars:nets} 

Prior to employing the mark correlation functions in Section  \ref{sec:global:chars} for studying the mark association/variation for marked point processes on linear networks, one first needs to adapt those functions so that they take the geometry of the underlying network into account. Hence, distances between any two points on a linear network are measured via different choices within the class of regular distances \citep{rakshit2017second}, depending on the application. Below, we briefly review mark correlation functions for marked point processes on linear networks.

Let $\tf_f^{\LL}: \M \times \M \to \R^+$ be a test function similar to those introduced in Section \ref{sec:global:chars}; the superscript $\LL$ is only used to emphasise the role of linear networks in this section.  For every pairs of points $(x, m(x)), (y, m(y)) \in X_{\LL}$, given that their interpoint network distance is $d_{\LL}(x, y)= r_{\LL}$, \cite{Eckardt:Moradi:currrent} defined 
    \begin{align*}
    c_{\tf_f^{\LL}}^{\LL}(r_{\LL})
    =
    \e 
    \left[
    \tf_f^{\LL} 
    \left(
    m(x), m(y)
    \right)
    \Bigl\vert 
    (x, m(x)), (y, m(y)) \in X_{\LL}
    \right],
\end{align*}
and $c_{\tf_f^{\LL}}^{\LL}=c_{\tf_f^{\LL}}^{\LL}(\infty)$.  
The $\tf_f^{\LL}$-correlation function is then similarly defined as
\[
\kappa^{\LL}_{\tf_f^{\LL}}(r_{\LL})
=
\frac{
c_{\tf_f^{\LL}}^{\LL}(r_{\LL})
}{
c_{\tf_f^{\LL}}^{\LL}
}.
\]
Next, we give examples of such mark correlation functions for network-constrained point processes with real-valued marks. By setting $m_1=m(x), m_2=m(y)$, the unnormalised mark differentiation function $\nabla_{mm}^{\LL}$  follows by specifying $\tf_f^{\LL}$ as $1-\min( m_1,m_2)/\max(m_1,m_2)$, yielding
\begin{small}
\begin{align*}
\nabla_{mm}^{\LL}(r_{\LL})
=
\e
\left[
1
-
\frac{
\min(m_1,m_2)
}{
\max(m_1,m_2)
}
\Biggl\vert 
(x, m_1), (y, m_2) \in X_{\LL}
\right],
\end{align*}
\end{small}
given that $d_{\LL}(x,y)=r_{\LL}$. \cite{Eckardt:Moradi:currrent} performed a simulation study where they considered different scenarios for the mark distribution of points and showed that when ignoring the geometry of the underlying network, the mark correlation functions in Section \ref{sec:global:chars} are not able to detect the true mark associations/variations.

\subsubsection{Summary statistics for marked point processes with function-valued marks}\label{sec:global:chars:functional:nets}

Now, we consider a point process $X_{\LL}$ on $\LL\times\F(\mathcal{T})$ with points on a linear network $\LL$ and their corresponding function-valued marks on  $\F(\mathcal{T})$  with $\mathcal{T}=(a,b), -\infty\leq a\leq b\leq \infty$. Similar to Section \ref{sec:global:chars:functional}, here marks are of the form $f(x_i)(t): \mathcal{T}\subseteq \R \mapsto \R$.
Writing $\tf_f^{\LL}:\F(\mathcal{T})\times \F(\mathcal{T})\to \R^+$ to denote a generalised test function with arguments $f_1=f(x)(t)$ and $f_2=f(y)(t)$ for a fixed time point $t$, all mark correlation functions can be defined using pointwise specifications analogous to Section \ref{sec:global:chars:functional}.

Denoting the counterparts of \eqref{eq:ctffun} and \eqref{eq:ctffunind} as  $c^{\LL}_{\tf_f^{\LL}}(r_{\LL}, t)$ and $c^{\LL}_{\tf_f^{\LL}}(t)=c^{\LL}_{\tf_f^{\LL}}(\infty, t)$, we can define the pointwise $\tf_f^{\LL}$-correlation function $\kappa^{\LL}_{\tf_f^{\LL}}(r_{\LL}, t)$ by
\[
\kappa^{\LL}_{\tf_f^{\LL}}(r_{\LL}, t)
=
\frac{
c^{\LL}_{\tf_f^{\LL}}(r_{\LL}, t)
}{
c^{\LL}_{\tf_f^{\LL}}(t)
}.
\]
All mark correlation functions for marked point processes on linear networks with function-valued marks can be derived through integration with respect to time $t$ as 
\[
\kappa^{\LL}_{\tf_f^{\LL}}(r_{\LL})
=
\int_{\mathcal{T}} 
\kappa^{\LL}_{\tf_f^{\LL}}(r_{\LL}, t)
\de t.
\]
Further details can be found in \citet{Eckardt:Moradi:STAT}

\section{Local Indicators of Mark Association}\label{sec:locals}

Next, instead of looking at the average mark associations/variations among marks, which is what global mark correlation functions focus on, we propose local mark correlation (LIMA) functions to uncover significant local mark associations/variations. Recall that global mark correlation functions may often fail to uncover the true association/variation among marks as different mark behaviours may mask each other \citep{chaudhuri2023trend, moradi2023hierarchical}.  Following the same structure as in Section \ref{sec:pre}, we define our LIMA functions for marked point processes on planar spaces as well as on linear networks when points are labelled by either real-valued or function-valued marks. For every individual point, these local indicators are of great use to uncover its specific interrelations with the marks of other points within its surroundings as functions of distances $r, r_{\LL} \geq 0$. 

\subsection{Marked point processes in planar spaces}
 
\subsubsection{Local indicators for spatial point processes with  real-valued marks}\label{sec:local:reals:planar}
 
Let $X$ be a marked point process on $\R^d\times \R$ with $m_i=m(x_i)$ denoting the mark of its $i$-th point. 
Moreover, consider 
$\tf_{f, i}:\R \times \R \mapsto \R^+$ as a (local) test function.  For a fixed point $(x_i, m_i) \in X$ and all other points $(x_j, m_j) \in X\setminus (x_i, m_i)$ conditional on $d_{\R^d}(x_i, x_j) =r$, we then have 
\begin{align}\label{eq:localctf}
 c_{\tf_{f, i}}(r)
 =
 \e_{(x_i, m_i)} 
 \left[
 \tf_{f, i} (m_i, m_j)
\Bigl\vert 
  (x_j, m_j) \in X\setminus (x_i, m_i) 
 \right].
\end{align} 
Note that we can here, in fact, write $\tf_{f, i}=\tf_{f}$; however, to distinguish between local and global cases, we explicitly use the index $i$.
The above conditional expectation of the test function under mark independence 
is represented by $c_{\tf_{f, i}}=c_{\tf_{f, i}}(\infty)$ with the form    
\begin{eqnarray}\label{eq:ExpectedIndMarksPsi}
c_{\tf_{f, i}}
=
\int_{\R}
\int_{\R}
\tf_{f, i}(m_i, m_j) \nu(\de m_i) \nu(\de m_j).
\end{eqnarray}
Note that in both $c_{\tf_{f, i}}(r)$ and $c_{\tf_{f, i}}$, the point $(x_i, m_i) \in X$ is fixed as we are here interested in studying its mark association with other points in its surroundings. 
Normalising $c_{\tf_{f,i}}(r)$ by $c_{\tf_{f,i}}$ yields the local $\tf_{f,i}$-correlation function $\kappa_{\tf_{f,i}}(r)$, 
\begin{eqnarray}\label{eq:local:tfcorr}
\kappa_{\tf_{f, i}}(r)
=
\frac{
c_{\tf_{f, i}}(r)
}{
c_{\tf_{f, i}}
}. 
\end{eqnarray}
Any mark correlation function that is defined following the above principle/formulations will be called a LIMA function. Note that different choices of $\tf_{f, i}$ give rise to different local mark correlation functions. 
Recalling the presentation of $c_{\tf_f}(r)$ in \eqref{eq:ctf:2ndorderproducts:ratio},  $c_{\tf_{f, i}}(r)$ can be rewritten as
\begin{align}\label{eq:ctf:2ndorderproductslocal:ratio}
c_{\tf_{f, i}}(r)
=
\frac{
\varrho^{(2)}_{\tf_{f, i}}(r)
}{
\varrho_i^{(2)}(r) 
}   
\end{align}
where the numerator $\varrho^{(2)}_{\tf_{f, i}}$ is  the local $\tf_f$-second-order product density function of
\begin{align}\label{eq:productlocal}
&
\alpha^{(2)}_{\tf_{f, i}}(B_1, B_2)
\\
&
=
\e_{(x_i,m_i)}
\Bigg[
\sum_{
(x_j, m_j) \in X\setminus (x_i, m_i)
}
\hspace{-0.5cm}
\tf_{f,i} 
(
m_i, m_j
)
\1_{B_1} \{ x_i \}
\1_{B_2} \{ x_j \}
\Bigg], \nonumber
\end{align}
for any $B_1,\ B_2 \in \mathcal{B}(\R^d)$ and the denominator  $\varrho_i^{(2)}(r)$ is, heuristically, the probability of observing a point $x_i$ and any point $x_j$ in the distinct areas $B_1$ and $B_2$, i.e., the second-order product density function of the ground process where the first point is treated as fixed.
The index $(x_i, m_i)$ in the expectation emphasises that the expectation depends on the fixed point $(x_i, m_i)$.
Thus, $\kappa_{\tf_{f, i}}(r)$ can be interpreted as a ratio of two infinitesimal probabilities adjusted by the expectation of the local test function under mark independence.


Letting $d=2$, by focusing on the variation of marks, suitable LIMA functions include the local mark variogram $\gamma_{m_im_j}(r)$ with local test function
\begin{eqnarray}\label{eq:localvario}
\tf_{f,i}(m_i, m_j)
=
0.5 \Vert m_i-m_j\Vert^2,   
\end{eqnarray}
and the local mark differentiation function $\nabla_{m_im_j}(r)$ with local test function
\begin{eqnarray}
~
\tf_{f,i}(m_i, m_j)
=
1-\min(m_i,m_j)/\max(m_i,m_j).
\end{eqnarray}
Furthermore, the local association between the $i$-th mark $m_i$ and the marks $m_j$ of neighbouring points at spatial distances $r>0$ can be investigated using the local mark correlation functions $\kappa^{\mathrm{Sto}}_{m_im_j}(r)$ (Stoyan), $\kappa^{\mathrm{Bei}}_{m_im_j}(r)$ (Beisbart and Kerscher), and $\kappa^{\mathrm{Ish}}_{m_im_j}(r)$ (Isham) with the test functions specified as   
\begin{eqnarray}\label{eq:kappa:sto:planar:local}
\tf_{f,i}(m_i,m_j)
= 
m_i m_j,
\end{eqnarray}
\begin{eqnarray}
\tf_{f,i}(m_i,m_j)
= 
m_i + m_j,
\end{eqnarray}
and
\begin{eqnarray}\label{eq:localisham}
\tf_{f,i}(m_i,m_j)
= 
m_i (m_j - \mu_j),
\end{eqnarray} 
where $\mu_j$ is the mean of all such $m_j$, 
and the local function $I^{\mathrm{Sch}}_{m_i m_j}(r)$ (Schlather) with the test function
\begin{align}
\tf_{f,i}(m_i,m_j)
= 
m_i (m_j - \mu_j(\mathbf{r})). 
\end{align}
where $\mu_j(\mathbf{r})$ is the mean of all such $j$-th marks for which $d_{\R^2}(x_i, x_j)=r$.
Moreover, the local $\mathbf{r}$-mark correlation function $\kappa_{\bullet m_j}(r)$ has the test function
\begin{eqnarray}\label{eq:rmark:planar:local}
\tf_{\bullet,j} 
= 
m_j.
\end{eqnarray}
Lastly, the local $I$ functions $I^{\mathrm{Shi}}_{m_i m_j}(r)$ (Shimantani) is of the form 
\begin{eqnarray}
\tf_{f,i}(m_i, m_j) 
= 
m_i (m_j - \mu_j),
\end{eqnarray}
which coincides with the local version of the Isham mark correlation function, given in \eqref{eq:localisham}.

In the independent mark scenario, the expectations of the above local test function coincide with their expected value when $r\rightarrow \infty$. Hence, under mark independence, $\kappa_{\tf_{f, i}}(r)$ becomes one for all values of $r$, and any deviations from unity indicate the presence of mark structural dependence between the $i$-th point and its neighbouring points. 
To provide a few examples, the normalisation factor, i.e.,\ equation \eqref{eq:ExpectedIndMarksPsi},  for the local test function \eqref{eq:kappa:sto:planar:local} becomes
\begin{eqnarray*}
    c_{\tf_{f,i}}
    =
    m_i
    \int_{\R}
    m_j
    \nu(\de m_j) 
   =
   m_i \mu_j,
\end{eqnarray*}
and for the local test function \eqref{eq:localvario} becomes
\begin{align*}
 c_{\tf_{f,i}}
    &=
    \int_{\R}
    0.5 (m_i-m_j)^2
    \nu(\de m_j)
    \\
    &=
    0.5
    \Big[
   ( m_i - \mu_j)^2 + \sigma^2_j
    \Big],
\end{align*}
where $\sigma^2_j$ is the variance of all such $m_j$.

 \subsubsection{Local indicators for spatial point processes with  function-valued marks}\label{sec:local:fcts:planar}

Given a point process $X$ on $\R^d\times\F(\mathcal{T})$ and defining the local test function $\tf_{f, i}:\F(\mathcal{T}) \times \F(\mathcal{T}) \mapsto \R^+$, distinct local mark summary statistics can be constructed based on a pointwise formulation. To this end, for a fixed $t \in \mathcal{T}$, let $f_i=f(x_i)(t)$ denote the function-valued mark for a specific point $i$ and $f_j=f(x_j)(t)$ denote the function-valued mark for any alternative point in $X\setminus (x_i, f_i)$ for which we have $d_{\R^d}(x_i, x_j) = r $. Then, setting $c_{\tf_{f, i}}(r, t)$ and $c_{\tf_{f, i}}(t)=c_{\tf_f, i}(\infty, t)$ for the expectations of the local test function for any pair of points at interpoint spatial distance $r$, yields the pointwise local $\tf_f$-correlation function $\kappa_{\tf_{f, i}}(r,t)$, 
\begin{eqnarray}\label{eq:local:tf:kappa:fct:pointwise}
    \kappa_{\tf_{f, i}}(r,t)
    =
    \frac{
    c_{\tf_{f, i}}(r,t)
    }{
    c_{\tf_{f, i}}(t)
    }.
\end{eqnarray}
By doing so, we obtain pointwise mark correlation functions with respect to fixed time $t \in \mathcal{T}$. Hence, one can obtain 
\[
\kappa_{\tf_{f, i}}(r)
=
\int_{\mathcal{T}}  \kappa_{\tf_{f, i}}(r,t)\de t.
\]
For example, the unnormalised local mark variogram $\gamma_{f_if_j}(r)$ can be constructed by first computing the conditional expectation $c_{\tf_{f, i}}(r,t)$ when having the test function $0.5(f_i(t)-f_j(t))^2$ and then integrating $c_{\tf_f,i}(r,t)$ over $\mathcal{T}$.

\subsection{Marked point processes on linear networks}

\subsubsection{Local indicators for spatial point processes on linear networks with  real-valued marks}

This section is devoted to generalising Section \ref{sec:local:reals:planar} to network-constrained spaces. Let $X_{\LL}$ represent a marked point process on $\LL\times\R$ with real-valued marks for which marks of points $x_i \in \LL$  are denoted by $m_i=m(x)$.
Then, by fixing the point $(x_i, m_i) \in X_{\LL}$, one can, similar to \eqref{eq:localctf} and \eqref{eq:ExpectedIndMarksPsi}, define $c_{\tf^{\LL}_{f, i}}^{\LL}(r_{\LL}),\ r_{\LL} \geq 0,\ $ and $c_{\tf^{\LL}_{f, i}}^{\LL}=c_{\tf^{\LL}_{f, i}}^{\LL}(\infty)$ for the conditional expectations of the local test function $\tf^{\LL}_{f, i}:\R \times \R \to \R^+$. Consequently, distinct local mark correlation functions can be constructed, focusing on local average association, variation or autocorrelation. Note that, here, spatial distances are measured by a regular distance $d_{\LL}$. 

Association-related LIMA functions for marked point processes on linear networks include, the local mark correlation functions $\kappa_{m_im_j}^{\LL,\mathrm{Sto}}(r_{\LL})$ and $\kappa_{m_im_j}^{\LL,\mathrm{Bei}}(r_{\LL})$ with test functions $\tf^{\LL}_{f, i}(m_i, m_j) = m_i m_j$ and $\tf^{\LL}_{f, i}(m_i, m_j) =  m_i+m_j$, to name a few, and the local $\mathbf{r}_{\LL}$-correlation function $\kappa^{\LL}_{\bullet m_j}(r_{\LL})$ with $\tf^{\LL}_{f, i}(m_i,m_j)=m_j$. Furthermore, useful test functions investigating local variations of marks include $\tf^{\LL}_{f, i}(m_i, m_j) = 0.5\cdot(m_i-m_j)^2$ and $\tf^{\LL}_{f, i}(m_i, m_j) = 1-(\min(m_i, m_j)/\max(m_i, m_j))$  yielding the local mark variogram $\gamma_{m_i,m_j}^{\LL}(r_{\LL})$ and mark differentiation function $\nabla_{m_i,m_j}^{\LL}(r_{\LL})$, respectively.

  \subsubsection{Local indicators for spatial point processes on linear networks with  function-valued marks}

Lastly, by increasing the complexity of marks to situations where they are functions/curves, we define LIMA functions for marked point processes on linear networks with function-valued marks. For $X_{\LL}$ denote by $(x_i, f_i)=(x_i, f(x_i)(t))$ a fixed point, and let $(x_j, f_j)$ be any arbitrary point in $X_{\LL} \setminus (x_i, f_i)$ for which $d_{\LL}(x_i, x_j)= r_{\LL}$.
To adapt the above construction principle to the present setting, we write $\tf^{\LL}_{f, i}:\F(\mathcal{T})\times \F(\mathcal{T})\to \R^+$ to denote a (local) test function with arguments $f_i$ and $f_j$. Depending on the test function, let $c^{\LL}_{\tf^{\LL}_{f, i}}(r_{\LL},t)$ denote the pointwise conditional expectation of $\tf^{\LL}_{f,i}$ at distance $r_{\LL} \geq 0$ and fixed 
functional argument $t\in\mathcal{T}$. Likewise, we can define the pointwise conditional expectation of the local test function $c^{\LL}_{\tf^{\LL}_{f, i}}(t)=c^{\LL}_{\tf^{\LL}_{f, i}}(\infty,t)$ as 
\[
c^{\LL}_{\tf^{\LL}_{f, i}}(t)
=
\int_{\F(\mathcal{T})}
\int_{\F(\mathcal{T})}
\tf^{\LL}_{f, i}
(f_i,f_j)
\nu(\de f_i)
\nu(\de f_j),
\]
and the local pointwise network-constrained $\tf^{\LL}_{f, i}$-correlation function $\kappa^{\LL}_{\tf^{\LL}_{f, i}}(r_{\LL}, t)$ as
\begin{eqnarray}
\kappa^{\LL}_{\tf^{\LL}_{f, i}}(r_{\LL},t)
=
\frac{
c^{\LL}_{\tf^{\LL}_{f, i}}(r_{\LL},t)
}{
c^{\LL}_{\tf^{\LL}_{f, i}}(t)
}.
\end{eqnarray}
Then, different association/variation/autocorrelation-related LIMA functions follow from the integration of $\kappa^{\LL}_{\tf^{\LL}_{f, i}}(r_{\LL}, t)$ and $c^{\LL}_{\tf^{\LL}_{f, i}}(r_{\LL}, t)$ yielding 
\[
c^{\LL}_{\tf^{\LL}_{f, i}}(r_{\LL})
=
\int_{\mathcal{T}} 
c^{\LL}_{\tf^{\LL}_{f, i}}(r_{\LL}, t)\de t.
\]
and 
\[
\kappa^{\LL}_{\tf^{\LL}_{f, i}}(r_{\LL})
=
\int_{\mathcal{T}}
\kappa^{\LL}_{\tf^{\LL}_{f, i}}(r_{\LL}, t)\de t
\]
as the unnormalised and normalised integrated mark correlation functions. 

\subsection{Estimation}

Now, we focus on presenting non-parametric estimators for the LIMA functions we previously proposed. Since we consider spatial point processes on different state spaces, i.e., $\R^2$ and linear networks, and with different marks, i.e., real-valued and function-valued marks, we present the estimators of the LIMA functions following the same structure. 

\subsubsection{
Spatial point processes in planar spaces with real-valued marks
}\label{sec:est:real:planar}

In order to present estimators for our proposed LIMA functions, we make use of the representation of $c_{\tf_{f, i}}(r)$ as the ratio of two product density functions as given in \eqref{eq:ctf:2ndorderproducts:ratio} and \eqref{eq:ctf:2ndorderproductslocal:ratio}. In particular, we present unbiased estimators for the numerator and denominator of  \eqref{eq:local:tfcorr} as 
\begin{eqnarray}\label{eq:est1}\\
\widehat{\varrho^{(2)}_{\tf_{f, i}}}(r)
&=&
\frac{1}{2\pi r |W|}
\sum_{(x_j, m(x_j))\in X}
\tf_{f, i}
\Big(
m(x_i),m(x_j)
\Big)\nonumber
\\ 
&\times& 
\mathcal{K}
\Big(
d_{\R^2}(x_i,x_j) - r
\Big) \nonumber
\end{eqnarray}
and 
\begin{eqnarray}\label{eq:est2}
\\
\widehat{\varrho^{(2)}_i}(r)
&=&
\frac{1}{2\pi r |W|}
\sum_{(x_j, m(x_j)) \in X}
\mathcal{K}
\Big(
d_{\R^2}(x_i,x_j) - r
\Big)\nonumber
\end{eqnarray}
leading to
\begin{align}\label{eq:est3}
\widehat{c_{\tf_{f, i}}}(r) 
=
\frac{
\widehat{\varrho^{(2)}_{\tf_{f, i}}}(r)
}{
\widehat{\varrho^{(2)}_i}(r)
},
\end{align}
where $\mathcal{K}(\cdot)$ is a kernel function, 
and $|W|$ is the area of the observation window $W$. In both estimators \eqref{eq:est1} and \eqref{eq:est2}, an edge correction factor could be included, but as pointed out by \citet{Illian2008}, it might be ignored when both numerator and denominator are estimated using the same estimation principle. Applying the Campbell formula, it can be shown that both estimators \eqref{eq:est1} and \eqref{eq:est2} are unbiased estimators such that their ratio, given in \eqref{eq:est3}, results in a ratio-unbiased estimator for $c_{\tf_{f, i}}(r)$ \citep{Illian2008, Chiu2013}. Having proposed estimators for the unnormalised LIMA functions, we next focus on the normalisation factor, which represents the expectation of the test function under mark independence. The normalisation factor, given in the denominator of \eqref{eq:local:tfcorr}, is estimated as 
\[
\widehat{c_{\tf_{f, i}}}
=
\frac{1}{n-1} \sum_{j}
\tf_{f, i}
\Big(
m(x_i), m(x_j)
\Big).
\]
As a showcase, for $\tf_{f, i}(m(x_i), m(x_j)) = m(x_i) m(x_j)$, we have 
\[
\widehat{c_{\tf_{f, i}}}
=
\frac{1}{n-1} \sum_{j}
m(x_i) m(x_j)
=
\frac{m(x_i)}{n-1} \sum_{j}
m(x_j)
.
\]
Note that the focus here is on the association between the mark of the $i$-th point and the marks of all the other 
points.

\subsubsection{
Spatial point processes in planar spaces with function-valued marks
}

In a similar manner as in Section \ref{sec:est:real:planar},  $c_{\tf_{f, i}}(r,t)$ can be estimated through the ratio of the estimators for the pointwise local second-order product density functions $\varrho^{(2)}_{\tf_{f, i}}(r, t)$ and $\varrho^{(2)}_{i}(r)$ which are of the form 
\begin{eqnarray*}
  \widehat{\varrho^{(2)}_{\tf_{f, i}}}(r,t)
  &=&
  \frac{1}{2\pi r |W|}
  \sum_{(x_j, f(x_j)(t) )\in X}
  \hspace{-.2cm}
  \tf_{f, i}
  \Big(
  f(x_i)(t),f(x_j)(t)
  \Big)
  \\
  &\times&
  \mathcal{K}
  \Big(
  d_{\R^2}(x_i,x_j)-r
  \Big),
\end{eqnarray*}
and 
\begin{eqnarray*}
\widehat{\varrho^{(2)}_i}(r)
&=&
\frac{1}{2\pi r |W|}
\sum_{(x_j, f(x_j)(t) )\in X}
\mathcal{K}
\Big(
  d_{\R^2}(x_i,x_j)-r
\Big),
\end{eqnarray*}
respectively. Further, the normalisation factor $c_{\tf_{f, i}}(t)$ is estimated as
\[
\widehat{c_{\tf_{f, i}}}(t)
=
\frac{1}{n-1}
\sum_{j}
\tf_{f, i}
\Big(
  f(x_i)(t),f(x_j)(t)
\Big)
\]
yielding the pointwise estimator for the local $\tf_f$-correlation function
$\kappa_{\tf_{f, i}}(r,t)$ as 
\begin{align*}
\widehat{\kappa_{\tf_{f, i}}}(r,t)
=
\frac{
\widehat{c_{\tf_{f, i}}}(r,t)
}{
\widehat{c_{\tf_{f, i}}}(t)
}
=
\frac{
\widehat{\varrho^{(2)}_{\tf_{f, i}}}(r,t) 
/
\widehat{\varrho^{(2)}_{\tf_{f, i}}}(r)
}{
\widehat{c_{\tf_{f, i}}}(t)
}.
\end{align*}
From both estimators $\widehat{c_{\tf_{f, i}}}(t)$ and $\widehat{\kappa_{\tf_{f, i}}}(r,t)$, the integrated estimators follow from integration over $\mathcal{T}$ giving rise to
\[
\widehat{c_{\tf_f, i}}(r)
=
\int_{\mathcal{T}}
\widehat{c_{\tf_{f, i}}}(r,t)\de t
\]
and 
\[
\widehat{\kappa_{\tf_{f, i}}}(r)
=
\int_{\mathcal{T}}
\widehat{\kappa_{\tf_{f, i}}}(r,t)\de t.
\]
\subsubsection{
Spatial point processes on linear networks with real-valued marks
}
Following the same procedure as in Section \ref{sec:est:real:planar}, we have 
\begin{align*}
&
\widehat{c^{\LL}_{\tf^{\LL}_{f, i}}}(r)
\\
& =
\frac{
\sum_{j}
\tf^{\LL}_{f,i}(m(x_i),m(x_j))
\mathcal{K}(d_{\LL}(x_i,x_j)-r)
}{
\sum_{j}
\mathcal{K}(d_{\LL}(x_i,x_j)-r)
},
\end{align*}
and consequently
\begin{align*}
\widehat{\kappa_{\tf^{\LL}_{f, i}}^{\LL}}(r)
= 
\frac{
\widehat{c^{\LL}_{\tf^{\LL}_{f, i}}}(r)
}{
\widehat{c^{\LL}_{\tf^{\LL}_{f, i}}}
},
\end{align*}
where $\widehat{c^{\LL}_{\tf^{\LL}_{f, i}}}$ is of the form
\[
\widehat{c^{\LL}_{\tf^{\LL}_{f, i}}}
=
\frac{1}{n-1}
\sum_{j}
\tf^{\LL}_{f, i}
\Big(
m(x_i), m(x_j)
\Big).
\]

\subsubsection{
Spatial point processes on linear networks with function-value marks
}
Lastly, for spatial point processes on linear networks with function-valued marks, estimators of the local functions $c^{\LL}_{\tf^{\LL}_{f, i}}(r)$ and $\kappa^{\LL}_{\tf^{\LL}_{f, i}}(r)$ can be obtained by integrating the local pointwise estimators
\begin{align*}
&
\widehat{c^{\LL}_{\tf^{\LL}_{f, i}}}(r,t)
\\
&
=    
\frac{
\sum_{j}
\tf^{\LL}_{f,i}
\Big(
f(x_i)(t),f(x_j)(t)
\Big)
\mathcal{K}
\Big(
d_{\LL}(x_i, x_j)-r
\Big)
}{
\sum_{j}
\mathcal{K}
\Big(
d_{\LL}(x_i,x_j) - r
\Big)
},
\end{align*}
and 
\begin{eqnarray*}
\widehat{\kappa_{\tf^{\LL}_{f, i}}^{\LL}}(r,t)
=
\frac{
\widehat{c^{\LL}_{\tf^{\LL}_{f, i}}}(r,t)
}{
\widehat{c^{\LL}_{\tf^{\LL}_{f, i}}}(t)
}, 
\end{eqnarray*}
where 
\begin{eqnarray*}
\widehat{c^{\LL}_{\tf^{\LL}_{f, i}}}(t)
=
\frac{
\sum_{j}
\tf^{\LL}_{f,i}
\Big(
f(x_j)(t),f(x_j)(t)
\Big)
}{
n-1
}.
\end{eqnarray*}

\section{Simulation study}\label{sec:simu}

This section is devoted to evaluating the statistical performance of our proposed LIMA functions under different scenarios, comparing their performance to the existing
global mark correlation functions. More specifically, we focus on the probability of type I error and the power of the test. In other words, we investigate situations where no mark structure exists to determine how often global mark correlation functions and LIMA functions incorrectly identify mark structures, as well as cases where local mark structures are present in parts of the data to see how frequently global mark correlation functions fail to detect them, while LIMA functions successfully capture local mark behaviours.
To do so, we consider 500 realisations of a homogeneous Poisson point process with intensity function $500$ on a square-unit window, with four scenarios: 
I) marks are independently generated from a normal distribution $N(5, 0.5)$, II) two distinct areas as discs with radius $0.075$ exist, where in one of them, marks are independently generated from $N(7, 0.5)$, and in the other, from $N(3, 0.5)$, while marks outside these areas follow $N(5, 0.5)$; these areas contain approximately $4\%$ of the total points and their location randomly varies among $500$ simulated patterns, III) there are two areas of the same size as in scenario II where marks are generated from $N(7, 0.5)$, while points outside these regions are generated from $N(5, 0.5)$, and IV) points close to the diameter of the square-unit window have marks generated from $N(7, 0.5)$ while for the rest of the points marks are again generated from $N(5, 0.5)$. Regarding the spatial distribution of points, the same 500 unmarked generated patterns are used across different scenarios, allowing for a clearer comparison of the effects of different mark associations.
We use the \textsf{R} package \textsf{spatstat} \citep{Baddeley2015} and its sub-packages for simulations.

\begin{figure*}[t]
    \includegraphics[scale=0.08]{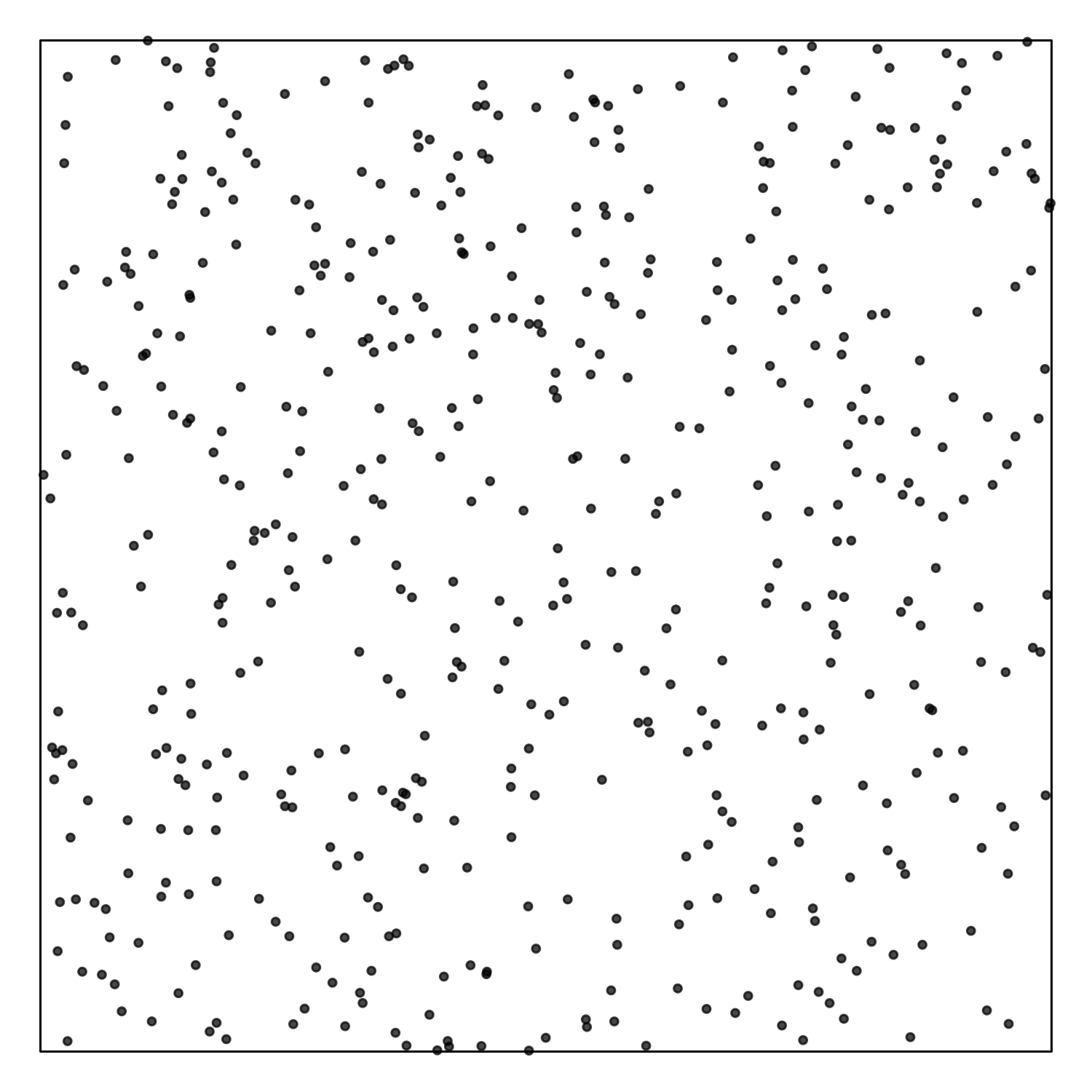}
    \includegraphics[scale=0.08]{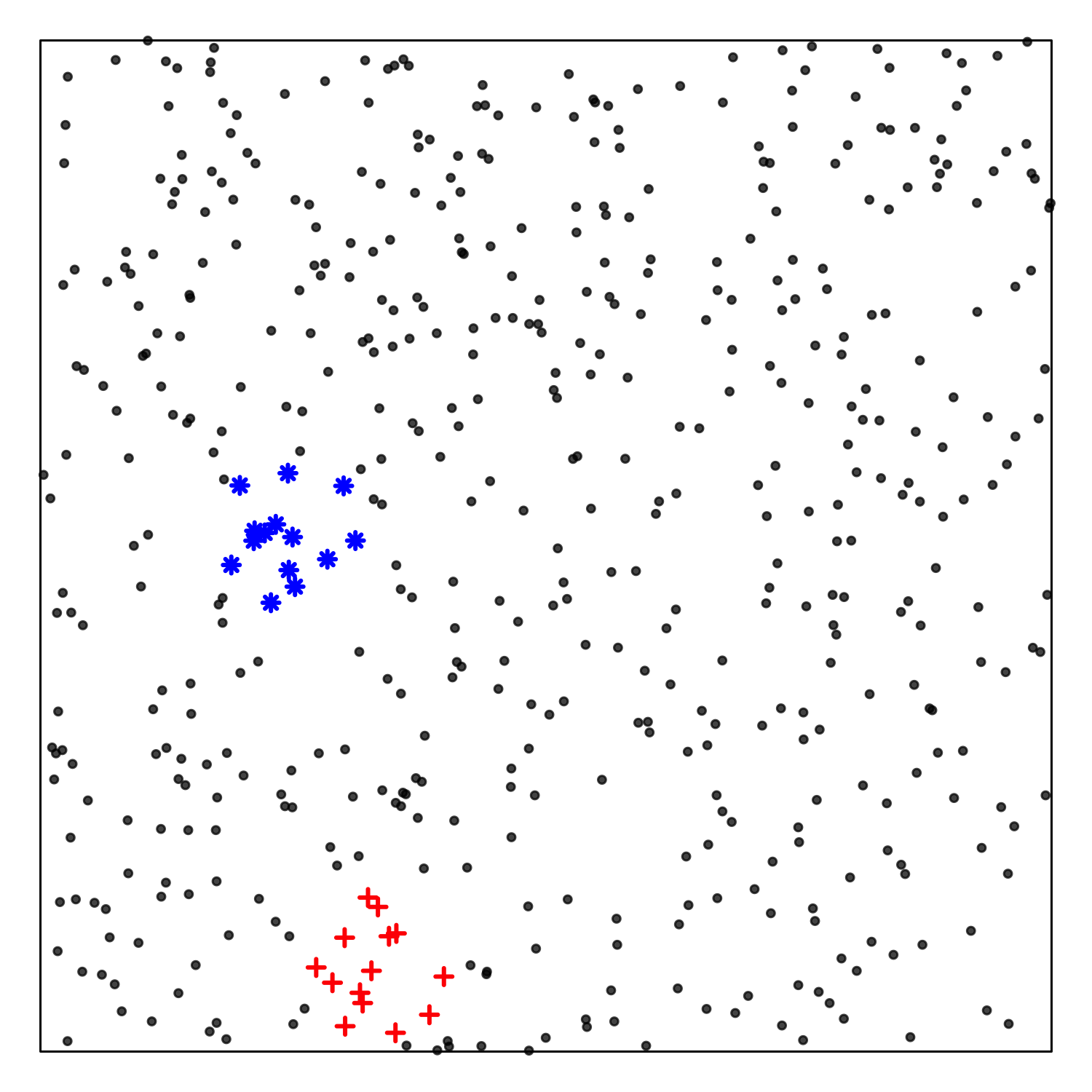}
    \includegraphics[scale=0.08]{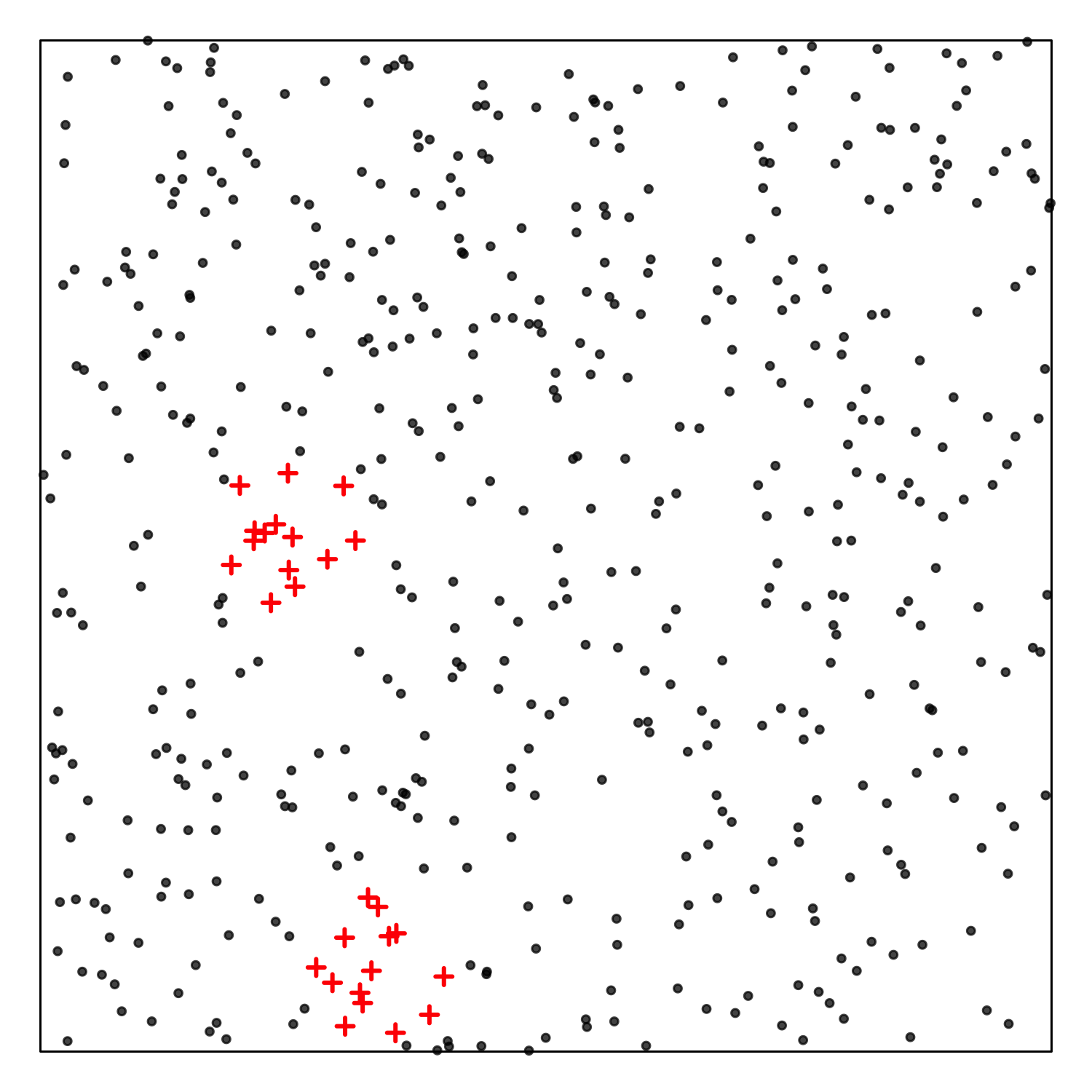}
    \includegraphics[scale=0.08]{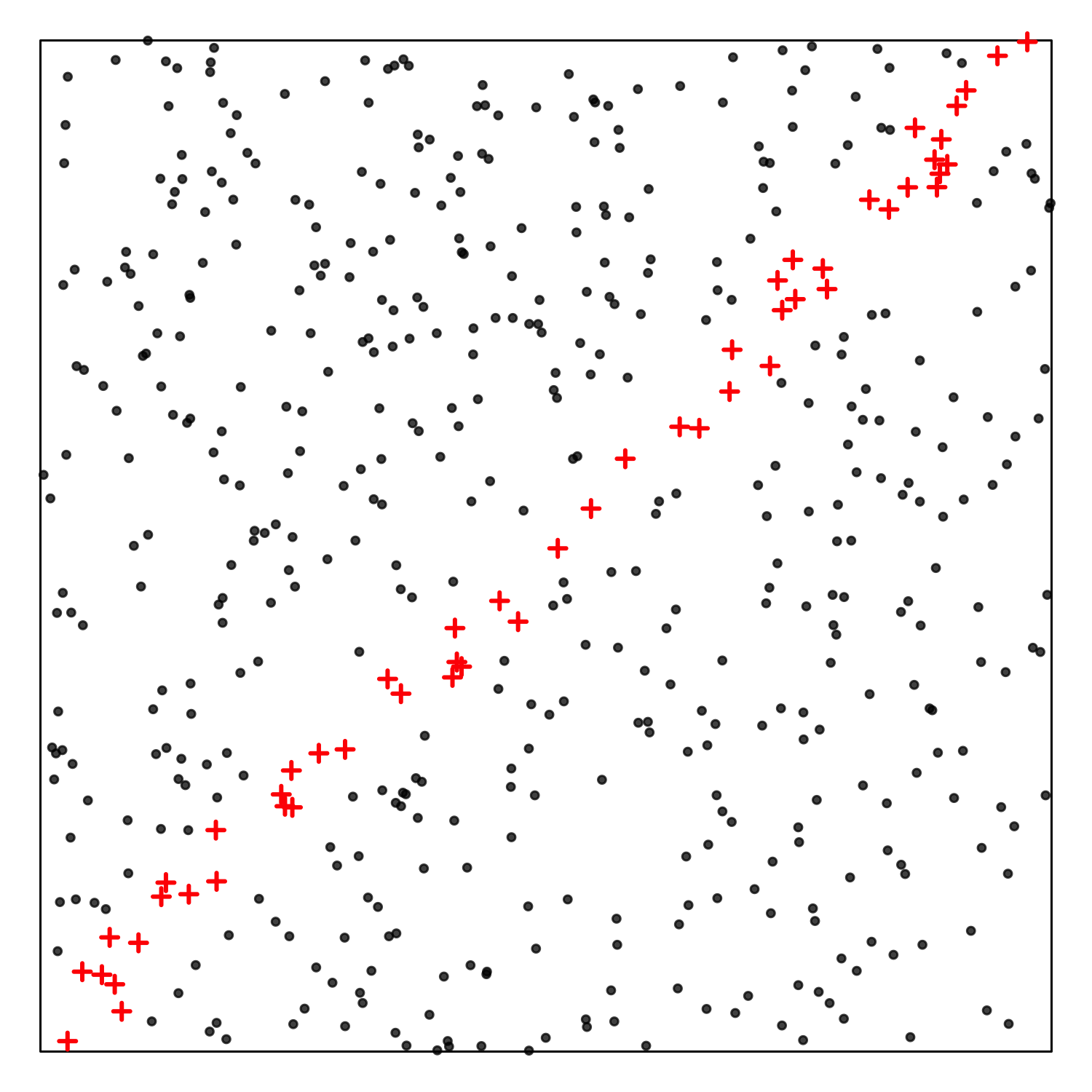}
    \caption{
    Examples of mark structures within the four scenarios.
    From left to right: scenario I, II, III and IV. For simplicity, mark values are ignored, and only points with different mark distributions are highlighted in different colours and shapes.
    }
    \label{Fig:example}
\end{figure*}

Within each scenario, for any single realisation, we use global envelope tests \citep{myllymaki2017global}, within the distance range $[0,r]=[0,0.25]$, to statistically identify any deviation from random labelling, i.e., verify the existence of any structure on the spatial distribution of marks. In the case of LIMA functions, we apply global envelope tests to each individual point. This allows us not only to detect deviations from random labelling but also to reveal which specific points contribute to the test's significance. Concerning the type of global envelop test, we employ the completely non-parametric rank envelope test based on extreme rank lengths, known as \textsf{`erl'} \citep{myllymaki2019get}, with 500 permutations. Figure \ref{Fig:example} shows one of the 500 generated patterns wherein, for each scenario, points within areas with different mark structures/associations are highlighted with different colours and shapes. Throughout the simulation study, we only use the local mark correlation functions $\kappa^{\mathrm{Sto}}_{m_im_j}$ as general results based on other LIMA functions will be similar. We also add that for some graphical representations, we make use of the \textsf{R} package \textsf{ggplot2} \citep{ggplot}.

\subsection{Scenario I}

\begin{figure}[h!]
    \centering
    \includegraphics[scale=0.08]{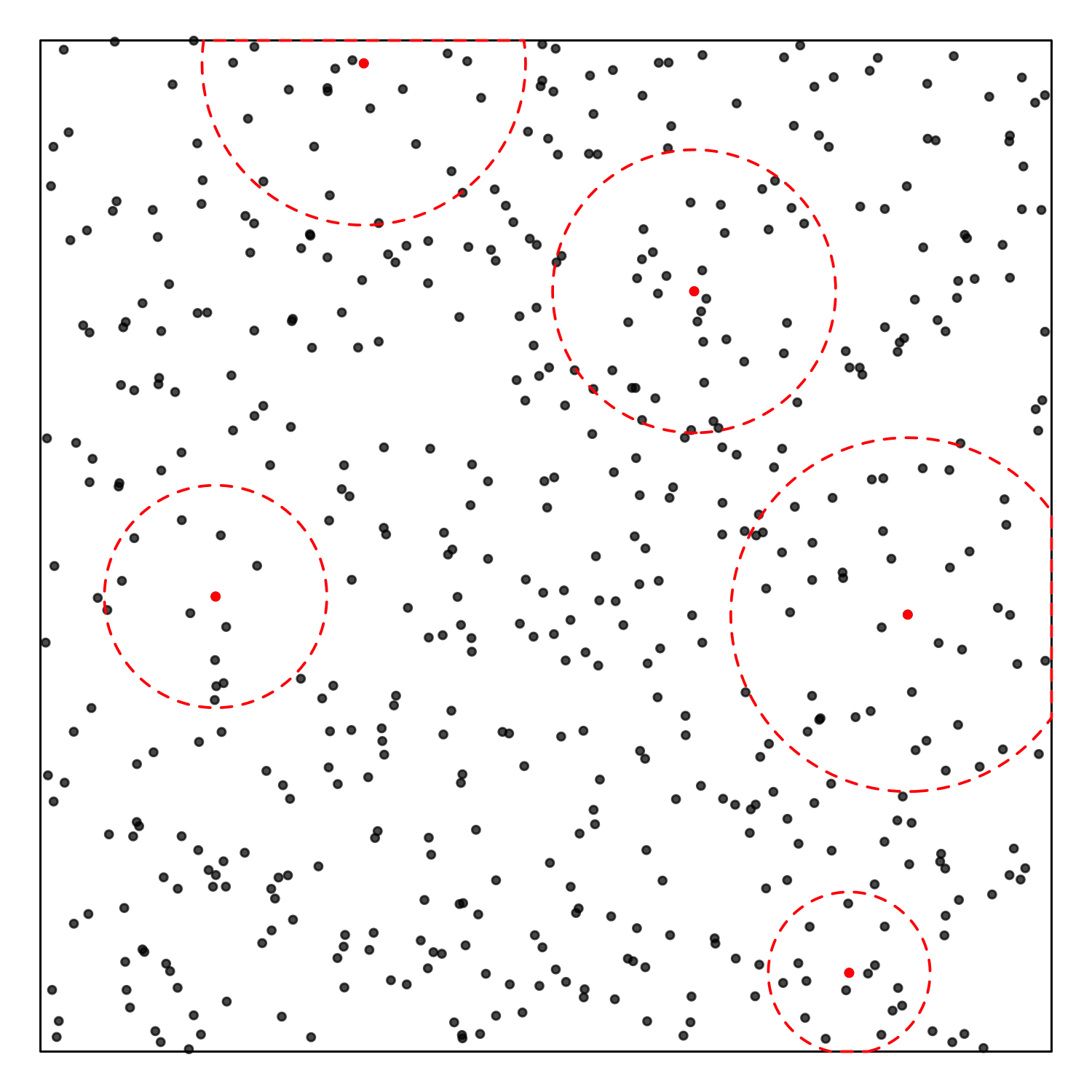}
    \includegraphics[scale=0.08]{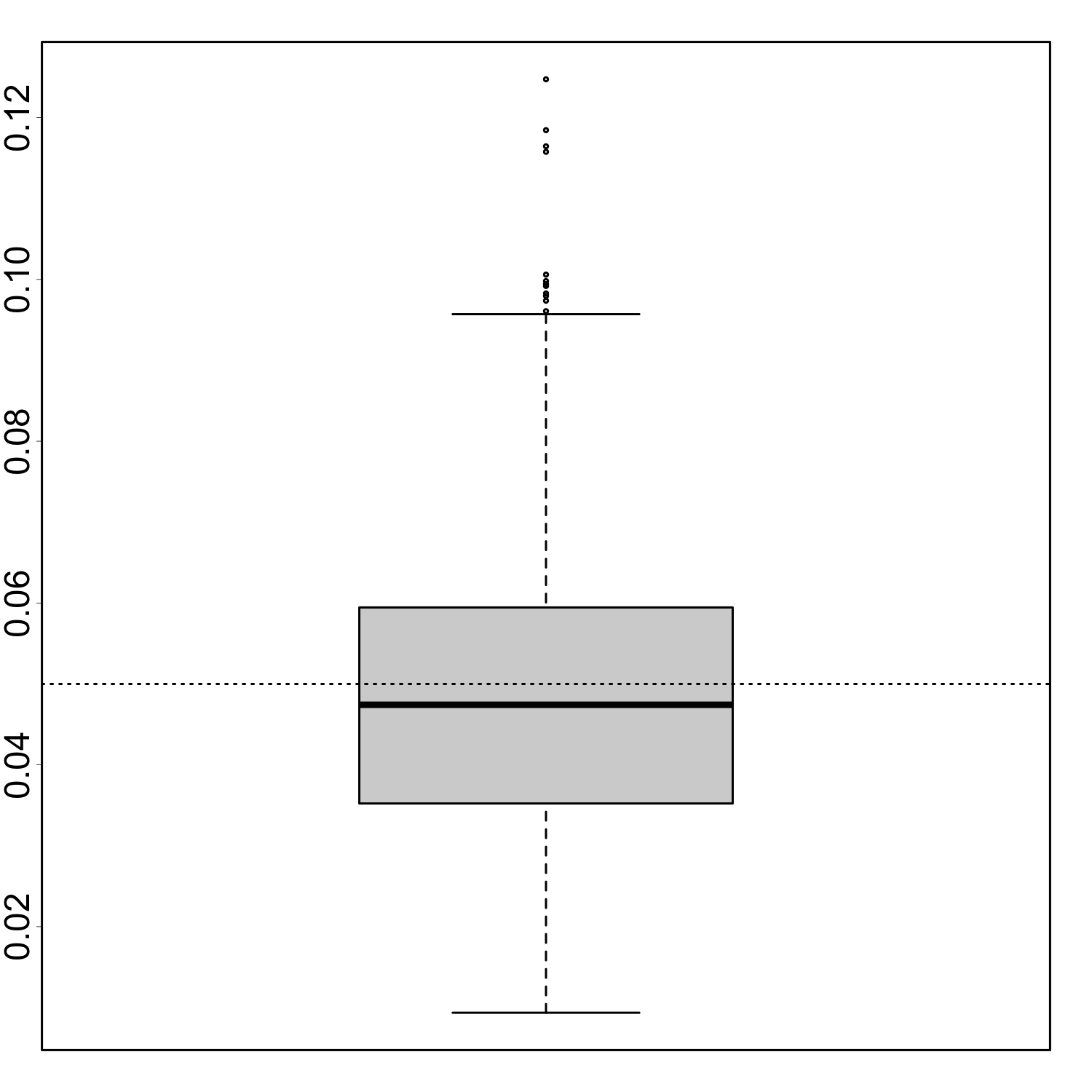}

    \caption{
   Left: One of the 500 simulated patterns, with significant points highlighted in red and their corresponding significant neighbourhoods shown as discs. 
   Right: Box-plot of the type I error probabilities across the 500 simulations, with the horizontal dashed line representing $0.05$.
    }
    \label{fig:s1out}
\end{figure}

As previously stated, in this scenario, the focus is on the performance of mark correlation functions when, in reality, there are no mark associations among points, and they are just randomly labelled by values generated from $N(5, 0.5)$. 
By applying global envelope tests with 500 permutations and a significance level of $0.05$, the global mark correlation function $\kappa^{\mathrm{Sto}}_{mm}$ mistakenly detects some mark structure/association among points in $6\%$ of the patterns, without providing any insight into which points contribute to the test's significance. For the local mark correlation function $\kappa^{\mathrm{Sto}}_{m_im_j}$, using the same procedure, we find that, on average, the contribution of individual points to the mark structure/association among points is mistakenly identified as significant for only $4.8\%$ of points. Note that by using our proposed LIMA functions, we get insight into individual points; this is not possible when using global functions. Thus, we are further interested in finding what makes those points significant. Points with significant LIMA functions for one of the $500$ simulated point patterns are depicted in the left plot in Figure \ref{fig:s1out} with corresponding discs highlighting the neighbourhood for which mark association deviates from random labelling. 
We observe that sparse neighbourhoods often surround these points. 
In other words, this might have been caused by the lack of enough information in their vicinity. Moreover, the right plot of Figure 
\ref{fig:s1out} shows the box-plot of all type I error probabilities of the LIMA functions for the $500$ simulated point patterns.

\subsection{Scenario II}

\begin{figure*}[t]
    \centering
    \includegraphics[scale=0.08]{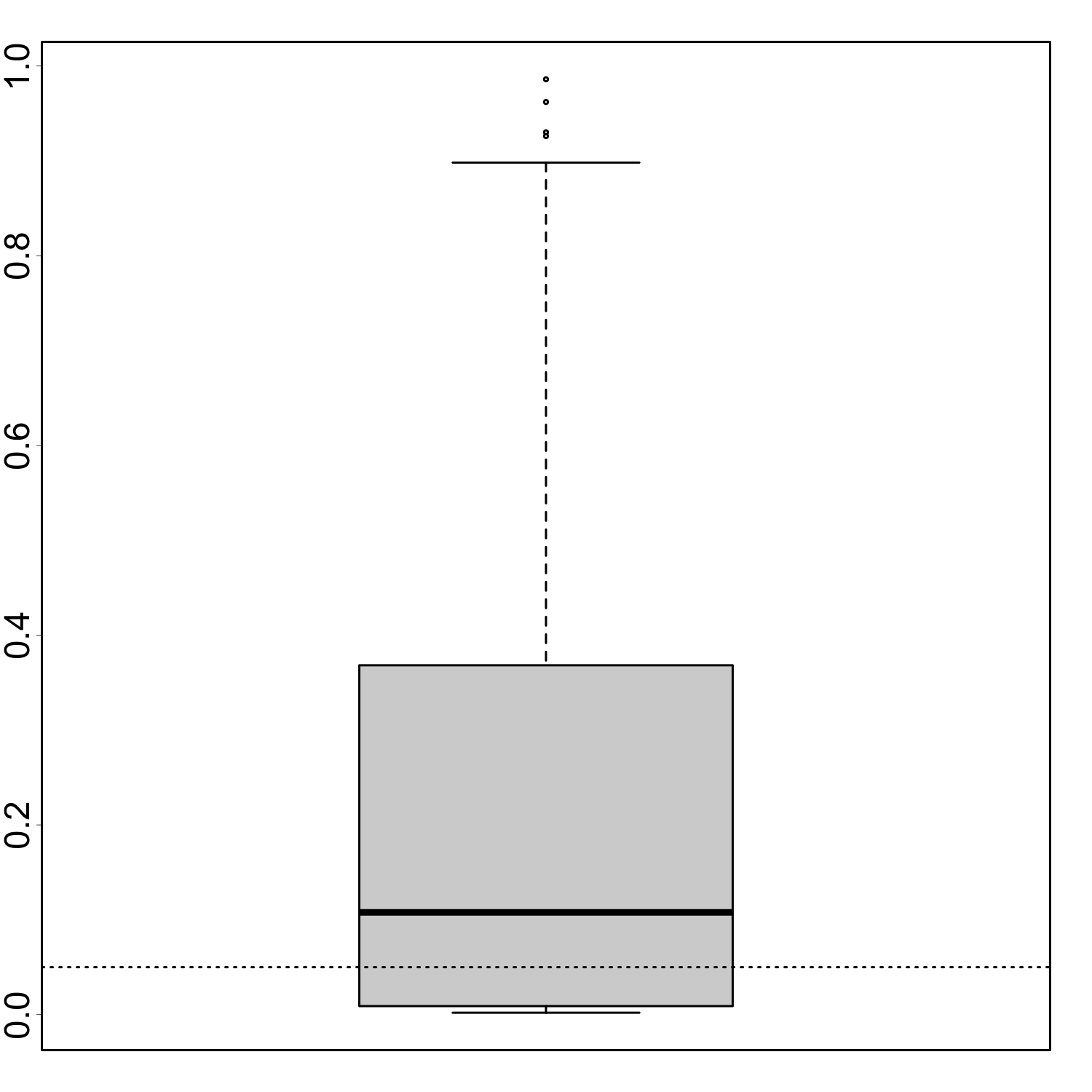}
    \includegraphics[scale=0.08]{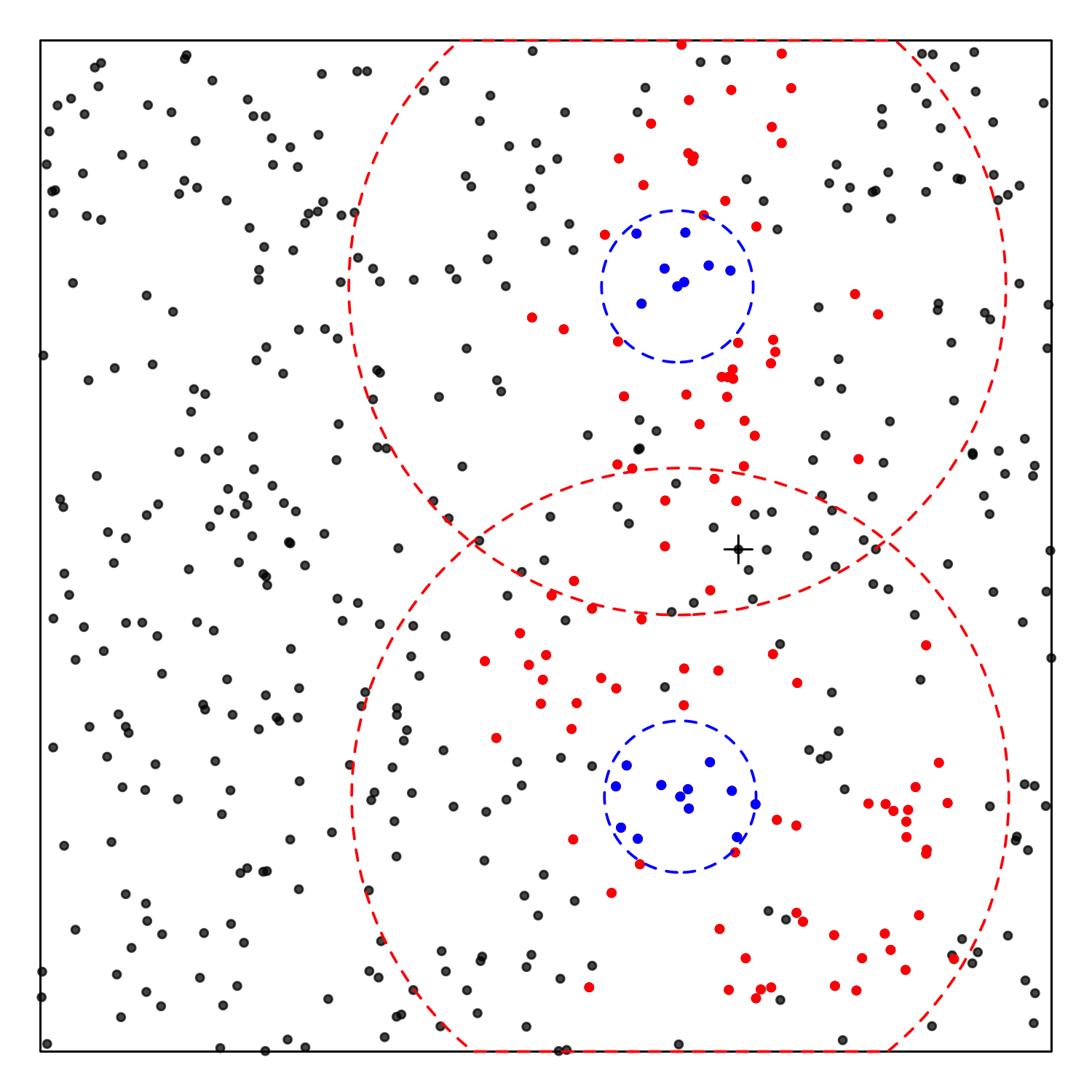}
    \includegraphics[scale=0.077]{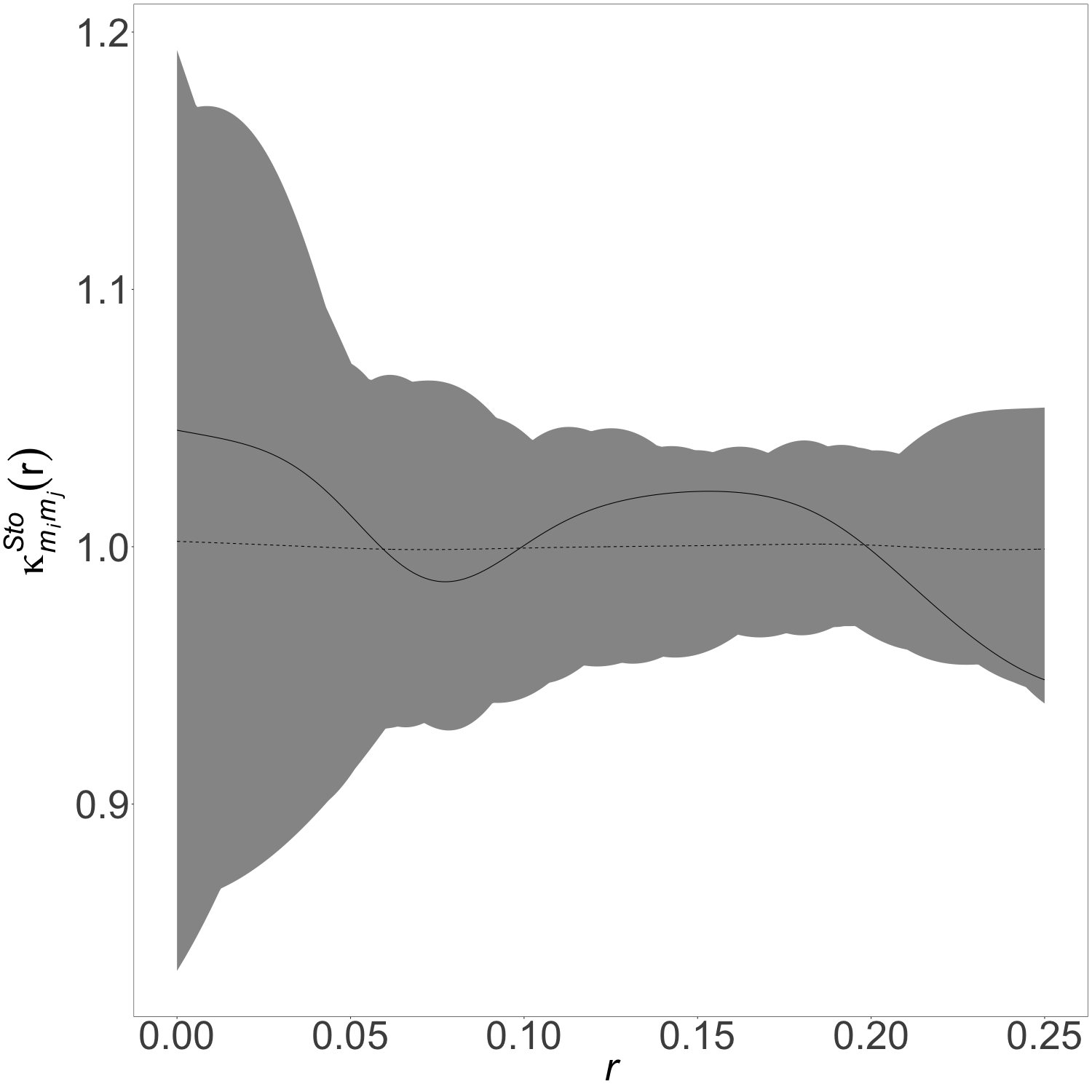}
    \caption{
    Results for scenario II.
    Left: Box-plot of $p$-values from $500$ global envelope tests using the global mark correlation function $\kappa^{\mathrm{Sto}}_{mm}$. Middle: One of the $500$ simulated point patterns, with regions of different mark distributions highlighted in blue, red discs of radius $r=0.25$ centred on these regions, and significant points marked in red or blue. A single point is highlighted as $+$. Right: Global envelope for the point highlighted as $+$ in the middle plot. The solid line is the local mark correlation function $\kappa^{\mathrm{Sto}}_{m_im_j}$ for the chosen point, and the dashed line shows the average of $\kappa^{\mathrm{Sto}}_{m_im_j}$ based on 500 permutations.
    }
    \label{fig:s2out}
\end{figure*}

This scenario represents a case where two regions exhibit distinct mark distributions compared to the rest of the observation window, with their mark behaviour potentially being masked by the mark distribution of points outside these regions. Based on the global envelope tests performed, the global mark correlation function $\kappa^{\mathrm{Sto}}_{mm}$ succeeds in only $40\%$ of the times to detect that there is some mark structure without offering additional insight into the nature of the structure. Moreover, Figure \ref{fig:s2out} shows the box-plot of the obtained $p$-values for the 
global envelope tests corresponding to the $500$ simulated patterns. However, the local mark correlation function $\kappa^{\mathrm{Sto}}_{m_im_j}$ successfully detects the mark structures for all $500$ patterns. It is important to note that with local mark correlation functions, not only are the points within the designated regions expected to be identified as significant, but also the points interacting with them, depending on the distance range $[0,r]$ used for the calculation. Thus, it is not immediate to calculate the power of the test for the local mark correlation function $\kappa^{\mathrm{Sto}}_{m_im_j}$ with respect to the number of tests performed. As an alternative, in  Figure \ref{fig:s2out}, we also show one of the simulated point patterns together with its points having significant local contributions highlighted in red/blue. The blue disc stands for the regions where the mark distribution differs from the rest of the points, and all points within these regions are correctly identified as having significant contributions. In addition, we also show red discs of radius $r=0.25$ (maximum $r$ considered) centred on these regions within which some of the points (red ones) are identified as significant, i.e., the red points, due to their interaction with the blue points. As shown in Figure \ref{fig:s2out}, no point beyond the distance  $r=0.25$ from the centre of the designed regions is identified as significant as they do not interact with blue points. In addition, we represent the global envelope test for a single point, highlighted as $+$ in the middle panel, for which the corresponding local mark correlation function stays within the envelope.

\subsection{Scenario III}

\begin{figure*}[t]
    \centering
    \includegraphics[scale=0.08]{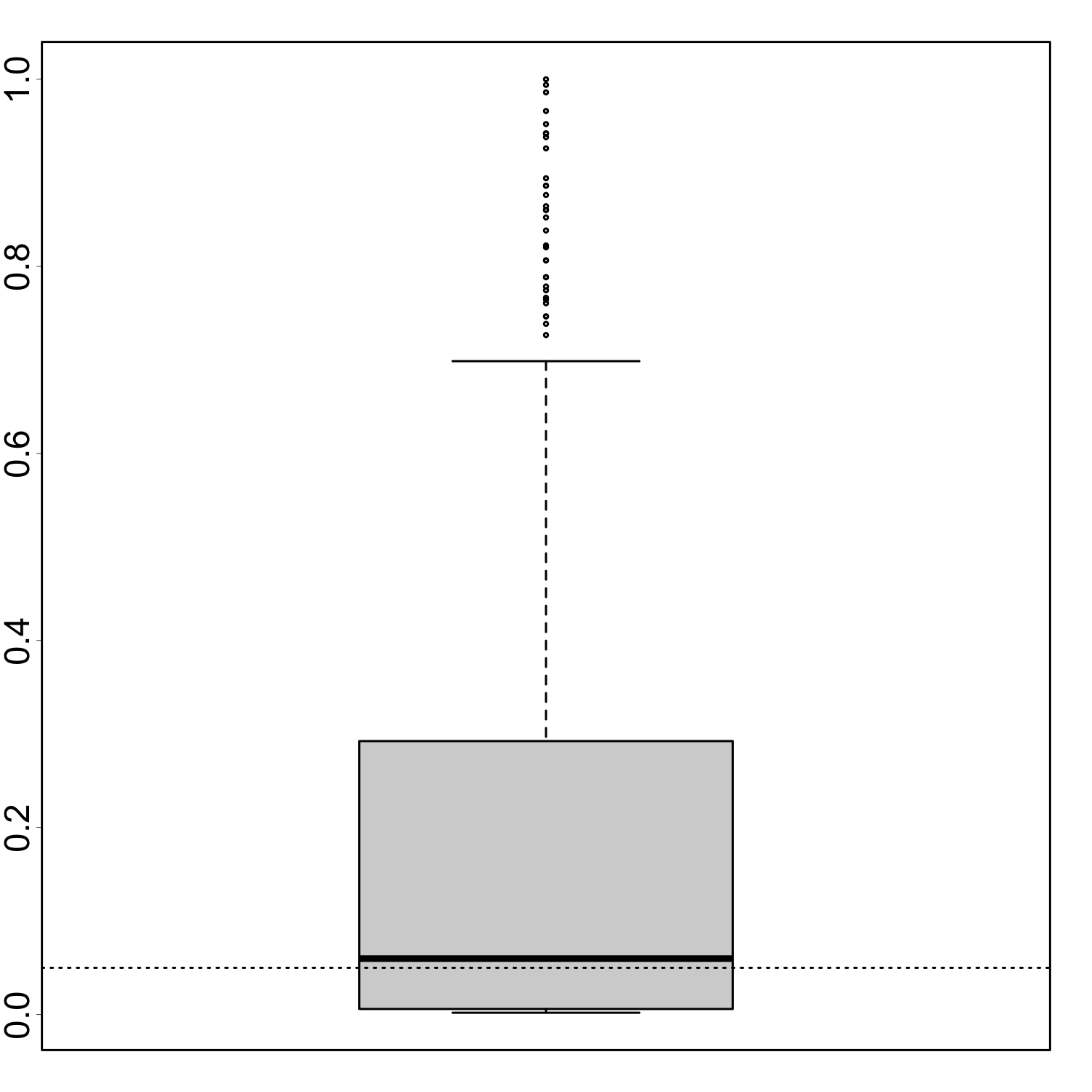}
    \includegraphics[scale=0.08]{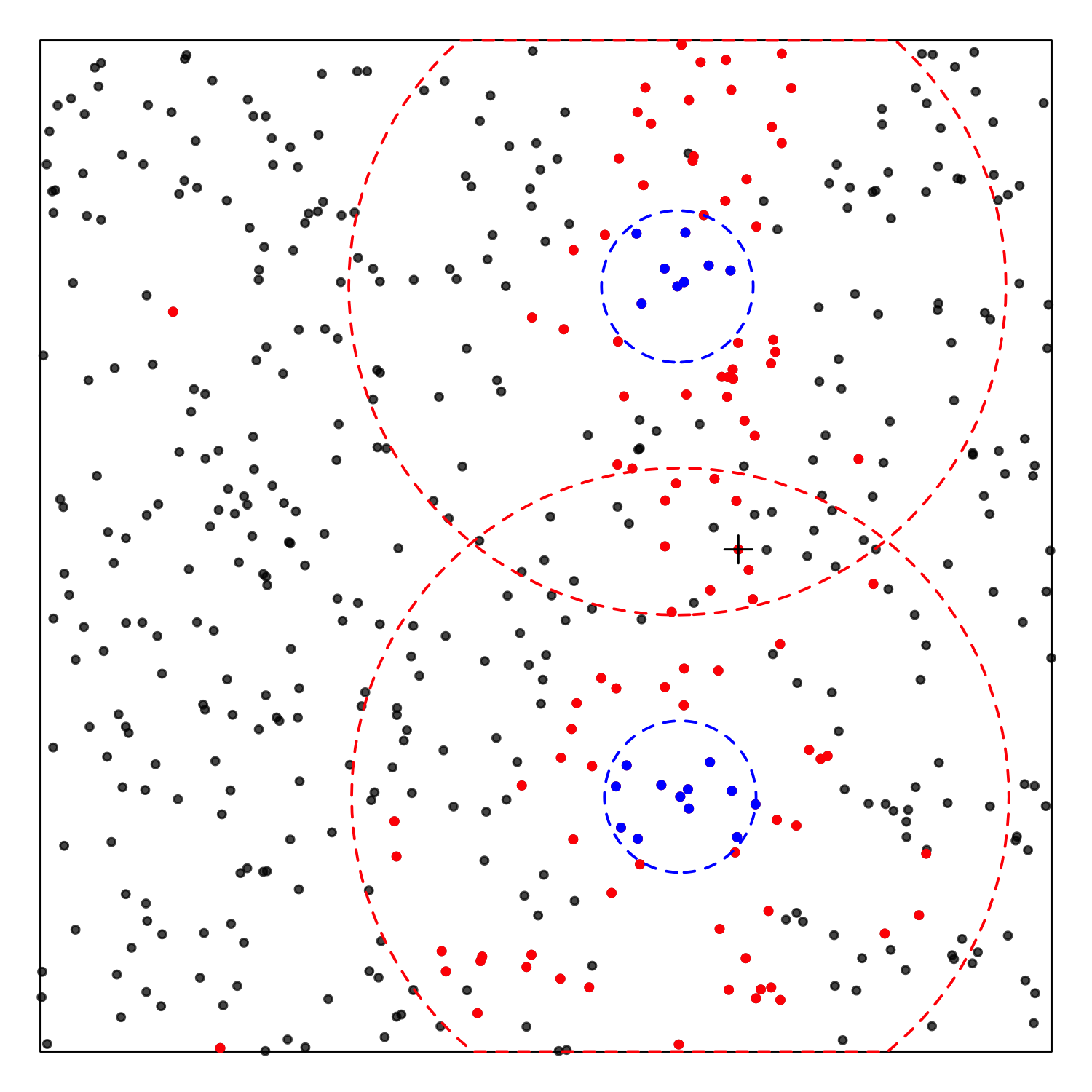}
    \includegraphics[scale=0.077]{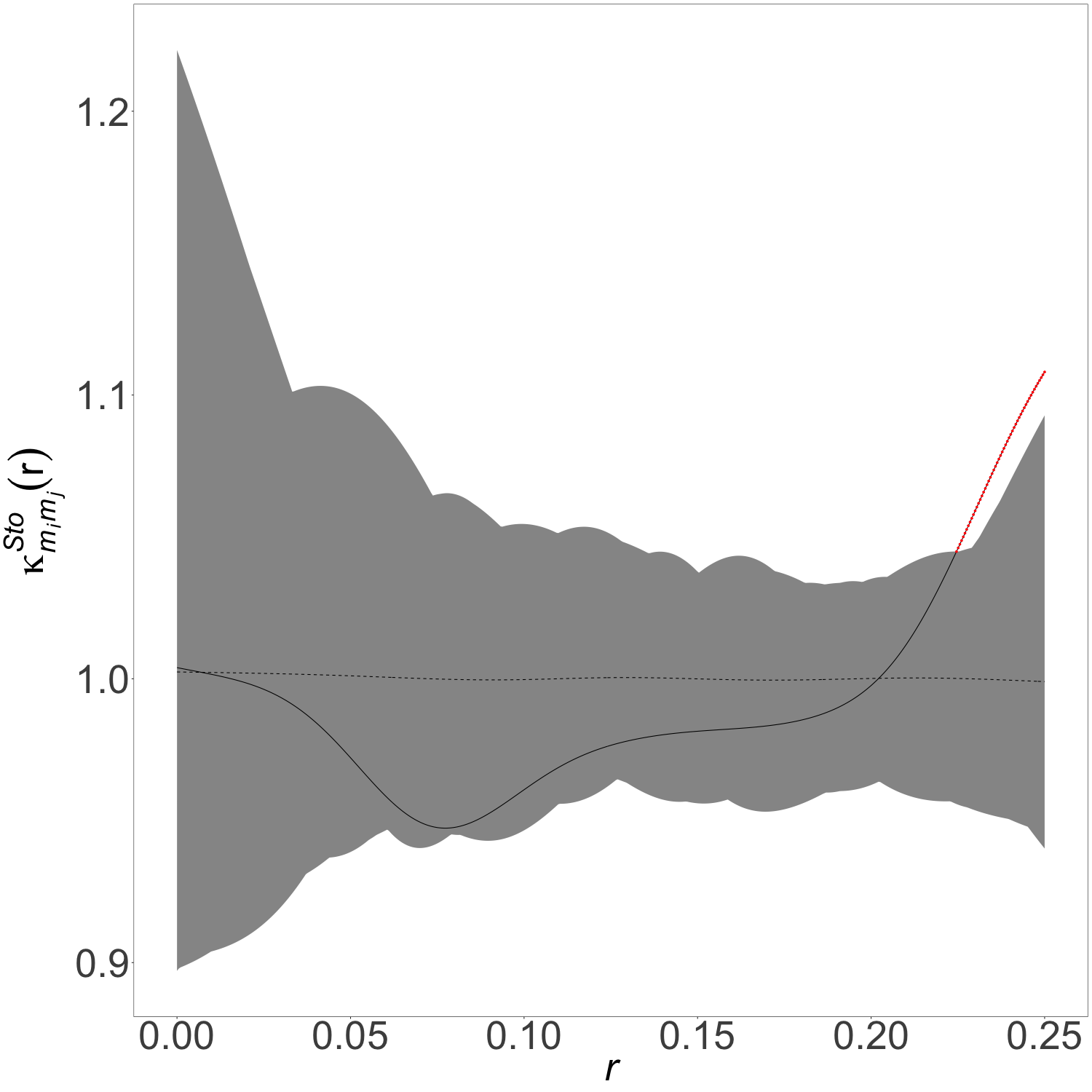}
    \caption{
    Results for scenario III. 
    Left: Box-plot of $p$-values from $500$ global envelope tests using the global mark correlation function $\kappa^{\mathrm{Sto}}_{mm}$.
    Middle: One of the $500$ simulated point patterns, with regions of different mark distributions highlighted in blue, red discs of radius $r=0.25$ centred on these regions, and significant points marked in red or blue.  A single point is highlighted as $+$.
    Right: Global envelope for the point highlighted as $+$ in the middle plot. The solid line is the local mark correlation function $\kappa^{\mathrm{Sto}}_{m_im_j}$ for the chosen point, and the dashed line shows the average of $\kappa^{\mathrm{Sto}}_{m_im_j}$ based on 500 permutations.
    }
    \label{fig:s3out}
\end{figure*}

This scenario is similar to the second one, except that the same mark distribution is applied within both considered regions. The use of the global mark correlation function $\kappa^{\mathrm{Sto}}_{mm}$ in combination with global envelope tests results in rejecting the hypothesis of random labelling only $47\%$ of the time, without indicating where the points with significant contributions are located. However, by employing the local mark correlation function $\kappa^{\mathrm{Sto}}_{m_im_j}$, in all $500$ simulated point patterns, local mark structures are correctly identified. Figure \ref{fig:s3out} shows the box-plot of the $p$-values for the $500$ global envelope tests based on the global mark correlation function $\kappa^{\mathrm{Sto}}_{mm}$, together with the results of the local mark correlation function $\kappa^{\mathrm{Sto}}_{m_im_j}$ for one of the simulated patterns; the same pattern as in Figure \ref{fig:s2out} is chosen. Similarly, we can see that all the points in the blue regions are identified as having significant contributions. In addition, some of the nearby points that interact with the points in the blue regions are also detected as significant. As for a comparison between scenarios II and III, we again focus on the same point as chosen in scenario II. Unlike scenario II, according to the global envelope tests, the contribution of this point is here identified as significant for distance $r\geq 0.22$. The difference in significance arises because, in scenario II, the point is surrounded by two regions with different mark distributions, which counterbalance each other's effects for distance $r\geq 0.22$. In scenario III, however, the point is surrounded by two regions with the same mark distribution, strengthening its interactions with nearby points, particularly within the distance range $[0.22, 0.25]$ as shown in the right panel of Figure \ref{fig:s3out}. However, when employing mark correlation functions and, in general, summary statistics for point processes, the focus is on smaller distances to better capture local behaviours, according to which, in both scenarios II and III, the local mark correlation function for this specific point stays within the envelope.

\subsection{Scenario IV}

\begin{figure}[!h]
    \centering
    \includegraphics[scale=0.08]{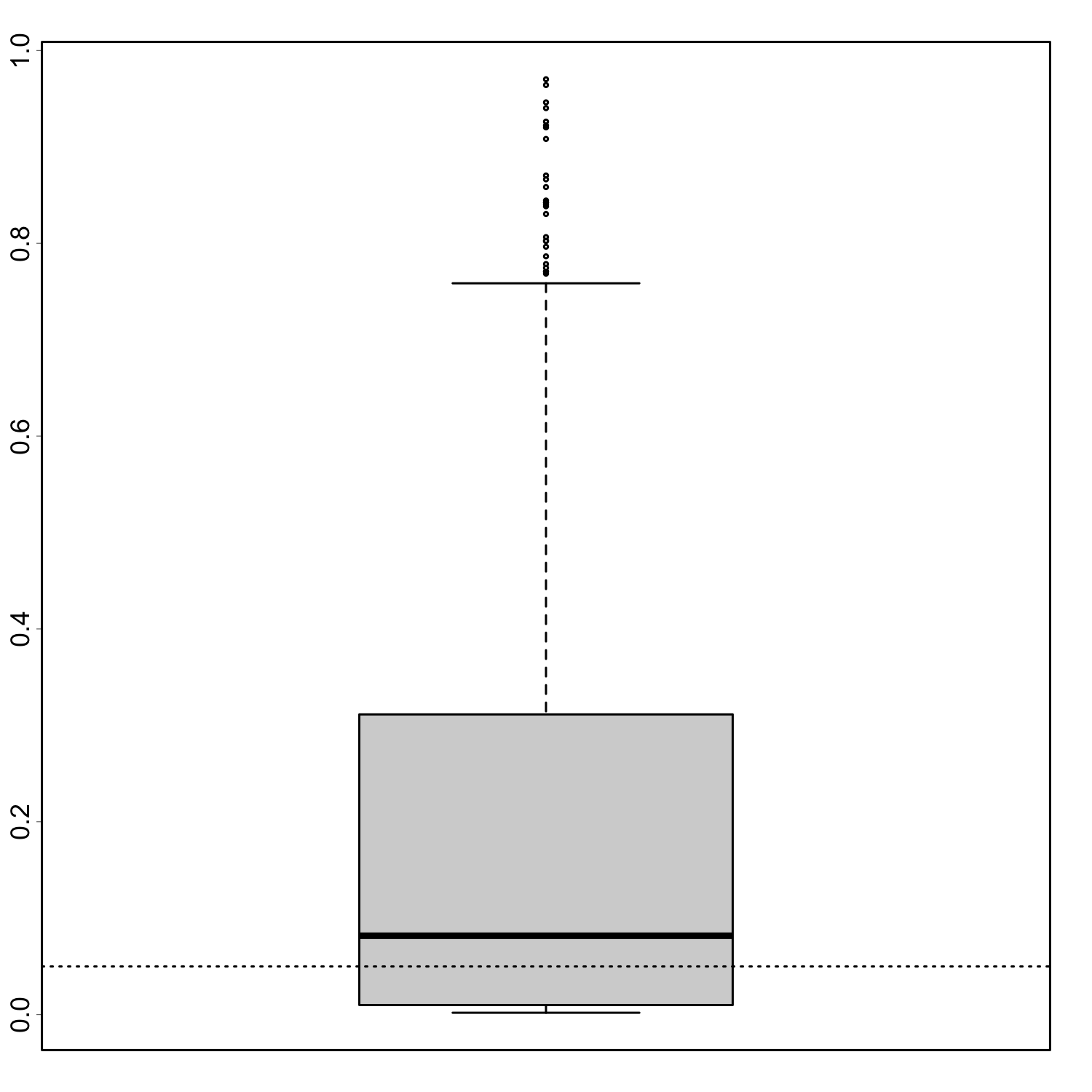}
    \includegraphics[scale=0.08]{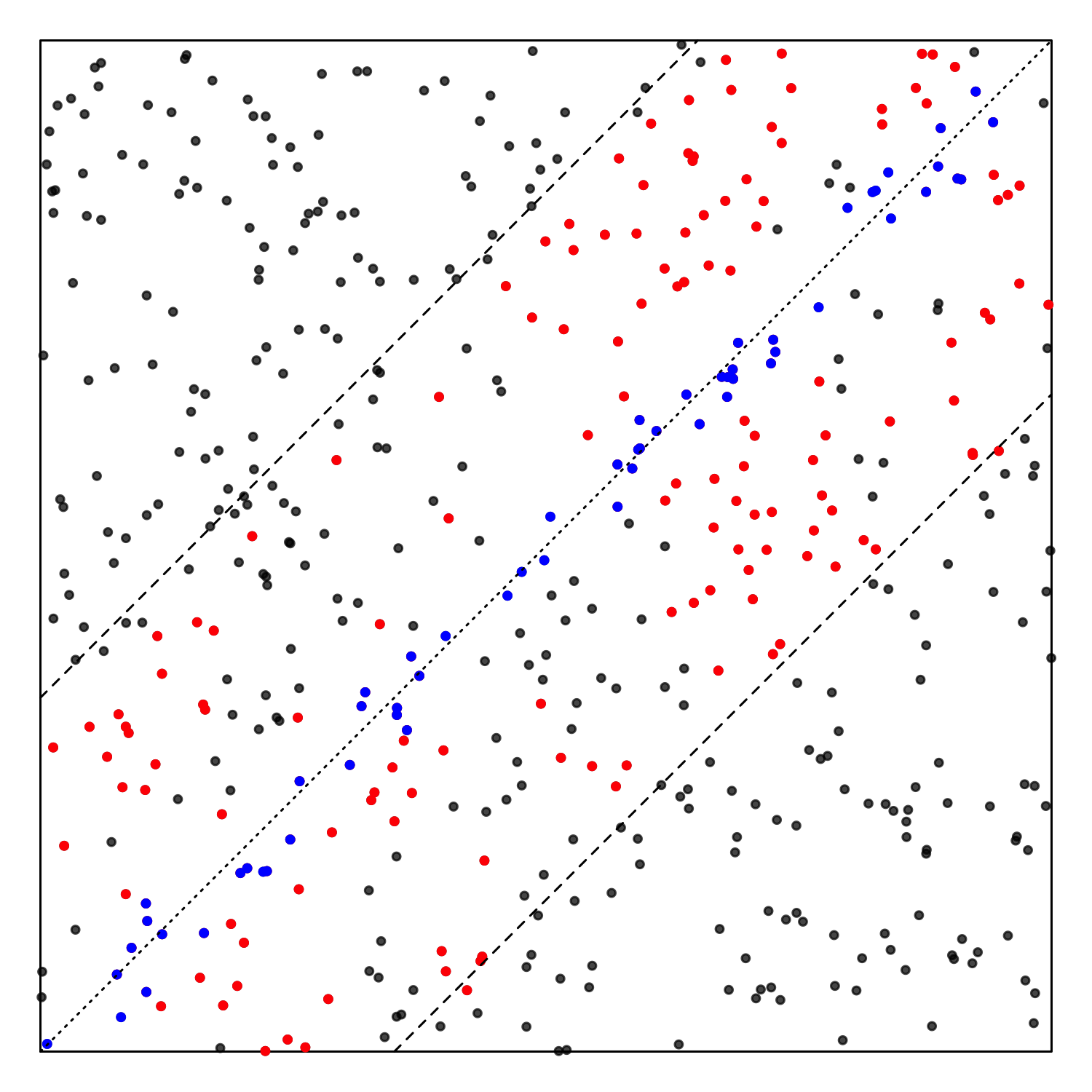}
    \caption{
    Results for scenario IV. 
    Left: Box-plot of $p$-values from $500$ global envelope tests using the global mark correlation function $\kappa^{\mathrm{Sto}}_{mm}$.
    Right: One of the $500$ simulated point patterns, with regions of different mark distributions highlighted in blue around a dotted line representing the window's diameter, significant points marked in red or blue, and two dashed lines parallel to the diameter, positioned $r=0.25$ units away. 
    }
    \label{fig:s4out}
\end{figure}

This scenario concerns the situation wherein points closer to the diameter of the observed window have larger marks. Combining the global mark correlation function $\kappa^{\mathrm{Sto}}_{mm}$ with global envelope tests based on 500 permutations, we find that in only $42.6\%$ of the times, the hypothesis of random labelling, at significance level $0.05$, has been rejected. Figure \ref{fig:s4out} shows the box-plot of the obtained $p$-values for the global envelope tests for the $500$ simulated patterns, having a third quartile of approximately $0.3$. Turning to our proposed LIMA  functions, we see that the local mark correlation function $\kappa^{\mathrm{Sto}}_{m_im_j}$ correctly detects some mark structure for all the $500$ simulated point patterns. As a showcase, we again revisit the specific pattern used in Figure \ref{fig:s2out} and \ref{fig:s3out}. From Figure \ref{fig:s4out}, we can see that all the points around the window's diameter with different mark distribution than the rest of the points are identified as significant. In addition to these points, some further points, having a distance less than   $r=0.25$ from the window's diameter, are also identified as significant due to their interactions with points around the window's diameter. It can be seen that no point beyond that distance is detected as significant since they have no local interaction with the blue points for $ r\leq 0.25 $.

\section{Application} \label{sec:apps}

This section focuses on applying our proposed local mark correlation functions within different applications dealing with distinct state spaces, such as $\R^2$ and linear networks with different types of marks, including real-valued and function-valued marks.

\subsection{Duke Forest data}

\begin{figure*}[t]
    \centering
    \includegraphics[scale=0.1]{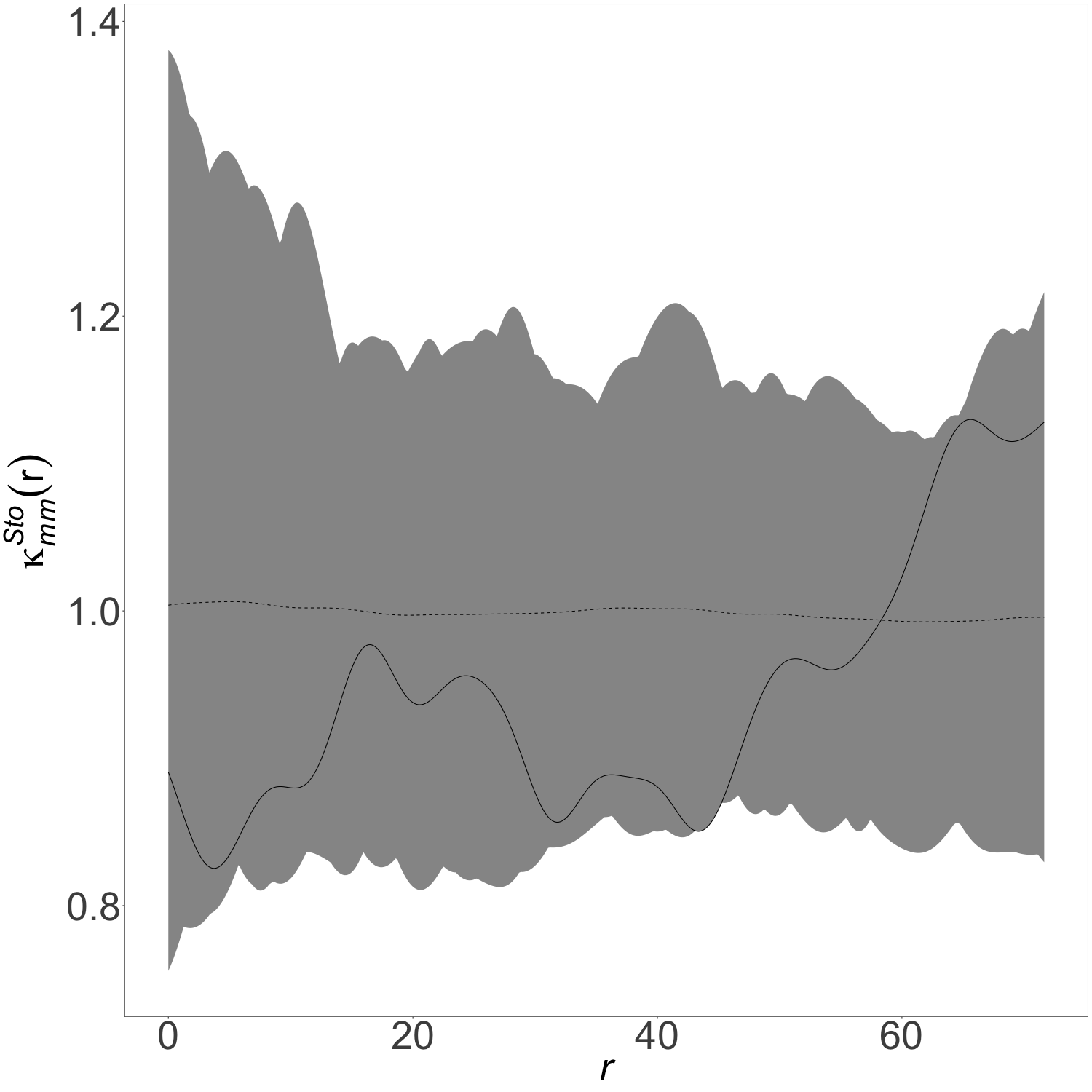}
    \includegraphics[scale=0.105]{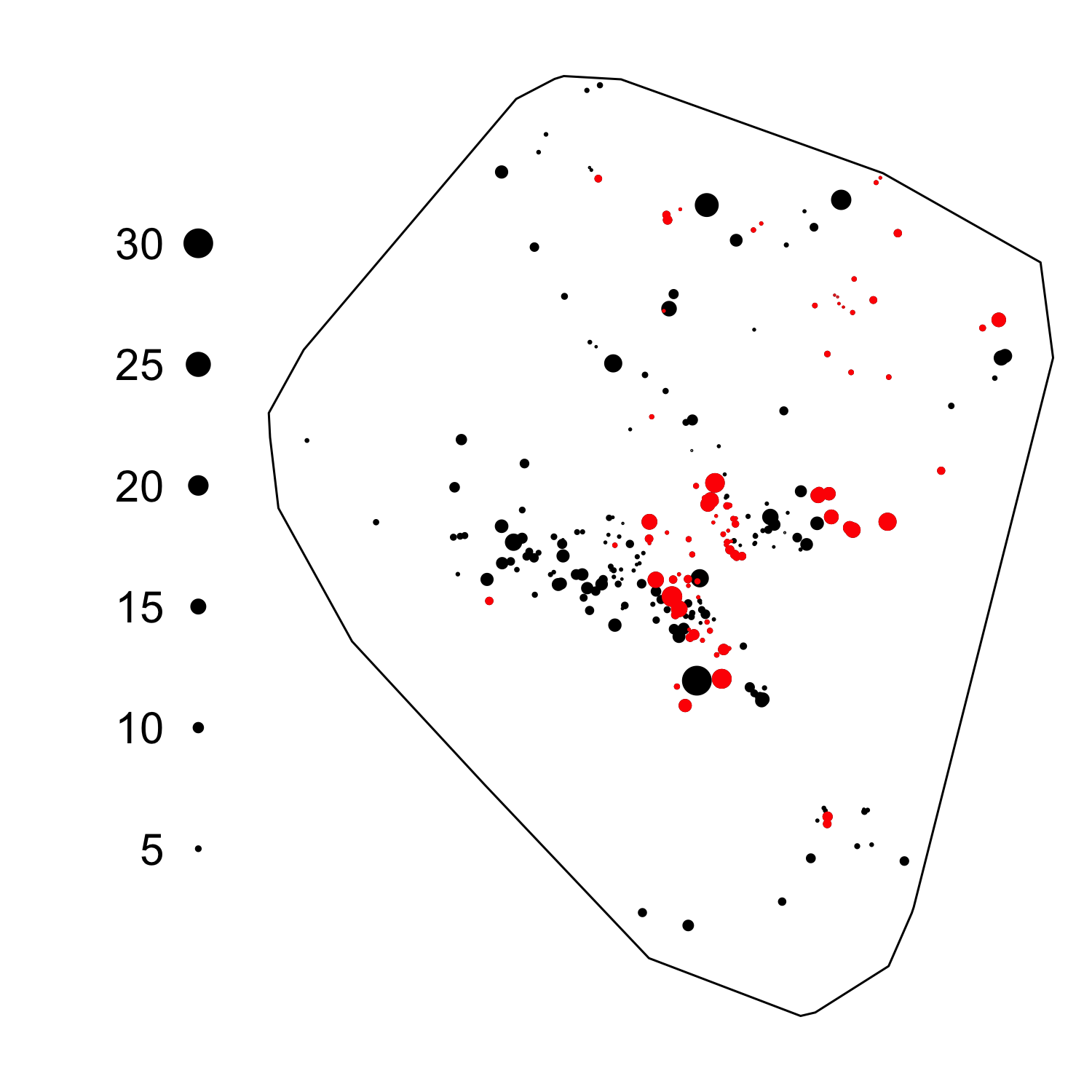}
    \includegraphics[scale=0.1]{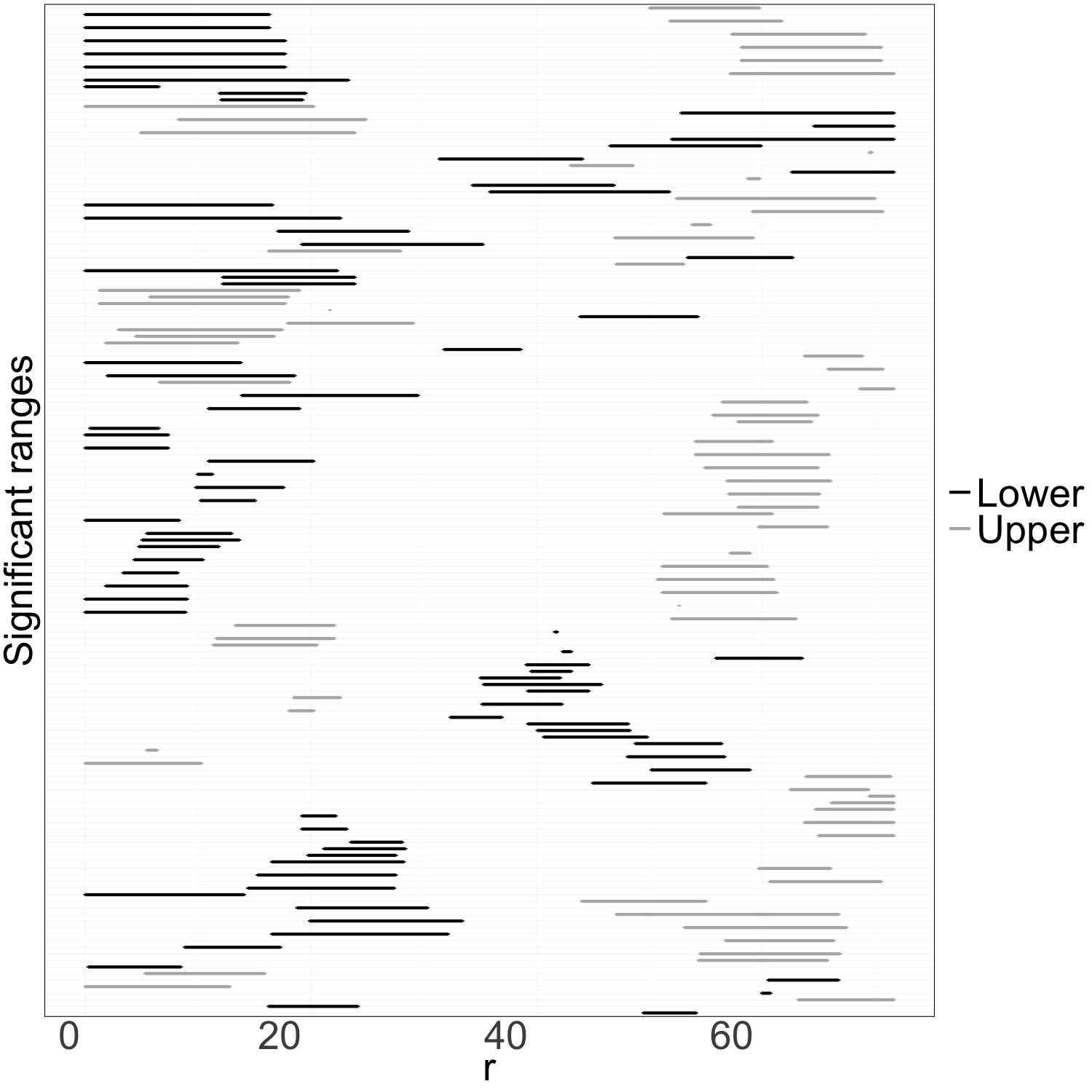}
    \caption{
    Results for Duke forest data.
    Left: Global envelope test using the global mark correlation function $\kappa^{\mathrm{Sto}}_{mm}$.
    Middle: The point pattern of the black gum trees with significant trees highlighted in red; the numbers show their corresponding dbh values.
    Right: Ranges of distances $r$ for which the local mark correlation function $\kappa^{\mathrm{Sto}}_{m_im_j}$ falls outside the envelope for the trees with significant mark associations represented in red in the middle plot. Lower and Upper refers to whether $\kappa^{\mathrm{Sto}}_{m_im_j}$ is falling outside the lower or upper bound.
    }
    \label{fig:duke}
\end{figure*}

The Duke forest data contains the locations of $10,053$ trees of $37$ species located in an area of size $65{\text{km}}^2$ (convex hull of the locations), which is split into three sub-areas: west, east, and south. The location of each tree is also labelled by its diameter at breast height (dbh), which was measured at some time during 2014. Most species are distributed within the three sub-areas. Here, we only focus on blackgum trees, which often have smooth leaves with non-toothed margins. The pattern of blackgum trees contains the locations of 276 trees, which can be seen in the middle plot given in Figure \ref{fig:duke} showing that blackgum trees are spread in the three sub-areas. The dbh of blackgum trees  (represented by the size of the points) varies between $0.7$ and $30.1$ with an average of $5.9$; the median is $4.35$.

In order to study the spatial association among marks and check any deviation from random labelling, we first make use of the global mark correlation function $\kappa^{\mathrm{Sto}}_{mm}$. 
The left panel of Figure \ref{fig:duke} shows the obtained global envelope based on 500 permutations under random labelling, wherein we can see that the estimated global mark correlation function $\kappa^{\mathrm{Sto}}_{mm}$ for the point pattern of blackgum trees stays within the envelope indicating no evidence against the assumption of random labelling, the corresponding $p$-value is $0.152$. Thereafter, we employ our proposed local mark correlation function $\kappa^{\mathrm{Sto}}_{m_im_j}$ for any single point out of which we found that at significance level $0.05$, $36.6\%$ of the points, corresponding to $101$ trees, are detected as significant which are highlighted in red in the middle panel of Figure \ref{fig:duke}. Recall that under the assumption of random labelling, we, in scenario I, showed that, on average, the probability of type I error for our local mark correlation functions is $0.048$. Additionally, it can be seen that significant points are generally located in the central part of the forest, with some tendencies toward the northeast. The dbh of these significant points varies between $0.7$ and $20$ with an average of $5.8$; the median is $4.2$. Furthermore, we add that out of the $101$ trees with significant local mark associations, for $41$ of them, the local mark correlation function $\kappa^{\mathrm{Sto}}_{m_im_j}$ stays outside the global envelope for some ranges within $r\leq 20$ meters revealing strong local associations in their surroundings. To provide further insights into this, for every significant point, we obtain the ranges of distances for which $\kappa^{\mathrm{Sto}}_{m_im_j}$, from either lower or upper bound, stays outside its corresponding envelope. These ranges are displayed in the right panel of Figure \ref{fig:duke}, showing that most significant points have significant mark associations within either small ($r \leq 20$) or large ($r \geq 60$) distances. Interestingly, those with significant mark associations within a distance of $r \leq 20$ fall outside the envelope from the lower bound, meaning that, for any such points, the product of their marks is smaller than that of random labelling. In other words, within the set of trees detected as significant, nearby trees have small dbh. In contrast, trees with significant mark associations within a distance of $r \geq 60$ fall outside the envelope from the upper bound, meaning that for any such pair of trees, at least one has a large dbh.

\subsection{Jersey City street crimes}

\begin{figure*}[!h]
    \centering
    \includegraphics[scale=0.1]{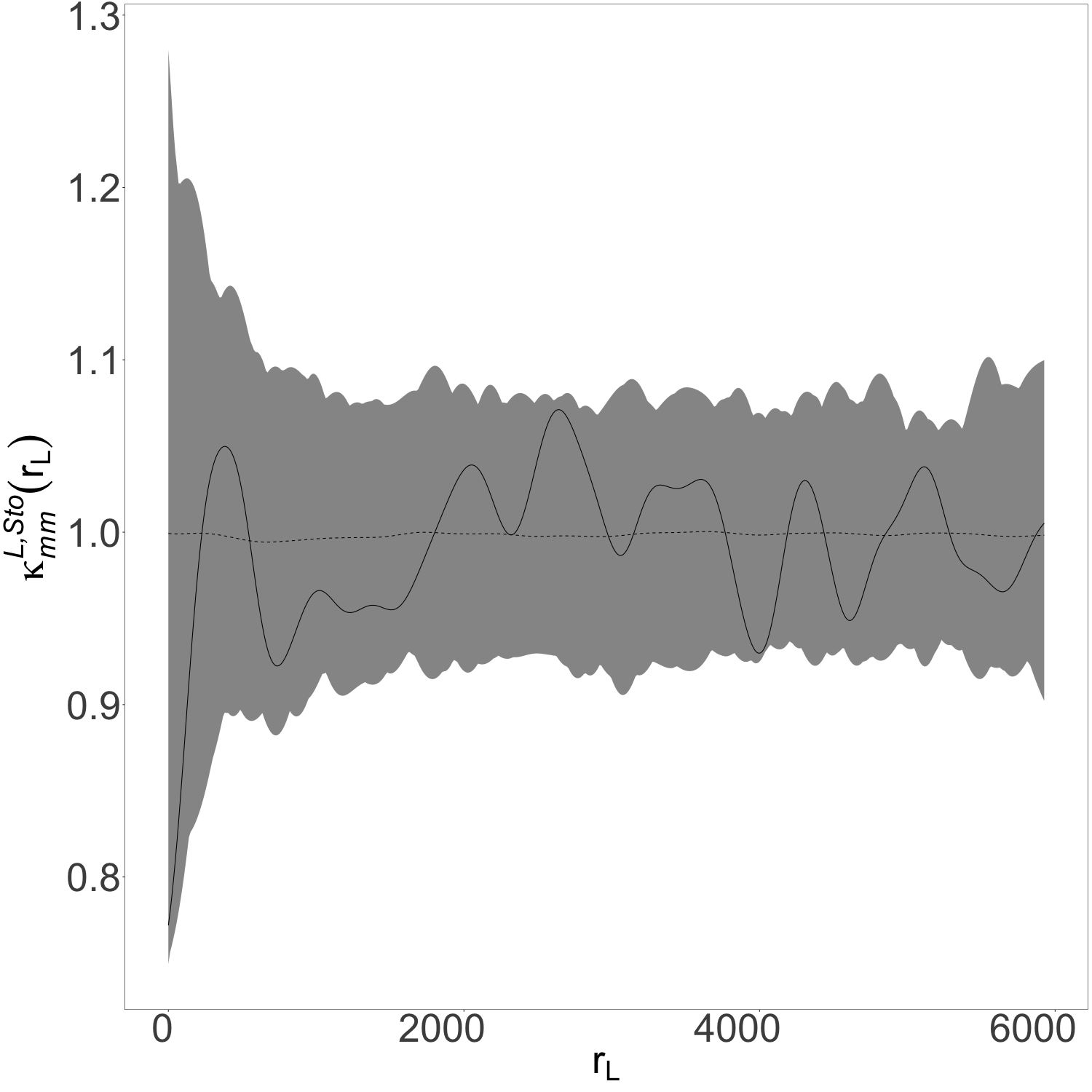}
    \includegraphics[scale=0.105]{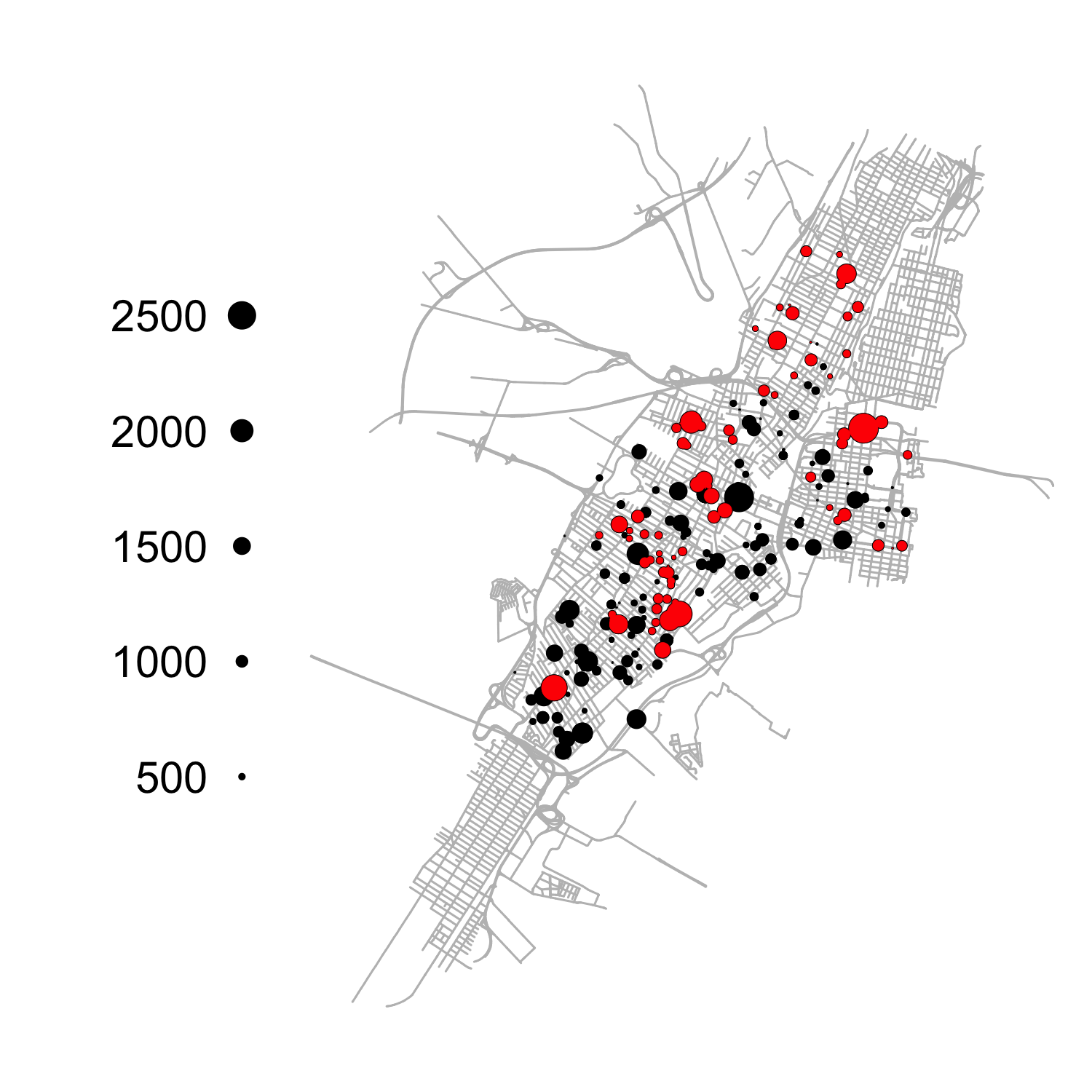}
    \includegraphics[scale=0.1]{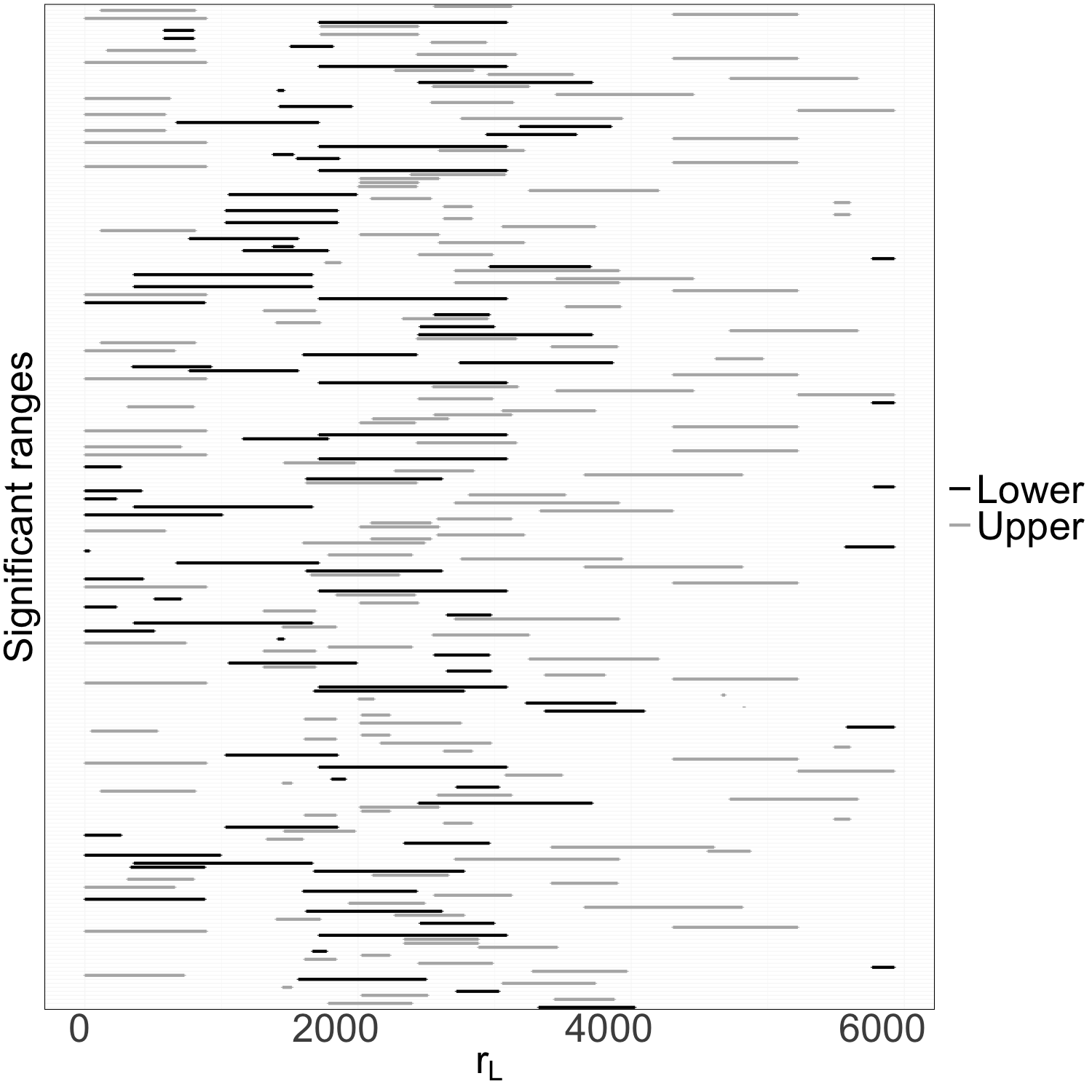}
    \caption{
     Results for Jersey City street crimes.
     Left: Global envelope test using the global mark correlation function $\kappa^{\LL,\mathrm{Sto}}_{mm}$.
     Middle: The point pattern of the crimes with significant ones highlighted in red; the numbers show their elapsed time in seconds. 
     Right: Ranges of distances $r_{\LL}$ for which the local mark correlation function $\kappa^{\LL, \mathrm{Sto}}_{m_im_j}$ falls outside the envelope for the crimes with significant mark associations represented in red in the middle plot. Lower and Upper refers to whether $\kappa^{\LL, \mathrm{Sto}}_{m_im_j}$ is falling outside the lower or upper bound.
    }
    \label{fig:jcr}
\end{figure*}

We now turn our attention to an application in which event locations are restricted to a linear network. Here, our objective is to analyse the time taken for the Jersey City Police Department to reach crime scenes; time elapsed refers to the time difference (in seconds) between when the police department receives a call and when they arrive at the crime scene. Data were accessed from the Jersey City data portal\footnote{\url{https://data.jerseycitynj.gov/pages/home-page/}}. The full dataset includes different crimes recorded during $2017$; some specific crimes, such as sexual assaults and attempted suicides, are not part of the full dataset. As a showcase, we here focus on crimes during March 2017, for which ambulances were requested. We excluded crimes for which elapsed time was exactly reported as zero. There is a total of $417$  street crimes on the street network of Jersey City, which has $24,936$ nodes and $26,824$ edges, with a total length of $961,328$ meters and a maximum node degree of $6$. The elapsed time has an average of $709.9$, a minimum of $60$, and a maximum of $2700$; the median is $600$. The middle panel of Figure \ref{fig:jcr} shows the locations of the crimes together with their elapsed time as marks.

We follow the same procedure in our simulation studies and the Duke Forest case. Initially, we employ the global mark correlation function $\kappa^{\LL,\mathrm{Sto}}_{mm}$ jointly with global envelope tests based on $500$ permutations. The left panel of Figure \ref{fig:jcr} shows that the global mark correlation function $\kappa^{\LL,\mathrm{Sto}}_{mm}$ stays within the envelope and finds no evidence of deviation from random labelling; the obtained $p$-value is $0.287$. Thereafter, we turn our attention to our proposed local mark correlation function $\kappa^{\LL,\mathrm{Sto}}_{m_im_j}$, which has the advantage of discovering locally significant contributions of individual points to the global mark correlation functions. According to the global envelope tests performed concerning each individual crime, based on $500$ permutations, $37.5\%$ of the crimes have significant local contributions within different distance ranges. The significant crimes are highlighted in red in the middle panel of Figure \ref{fig:jcr}, where most are located in the areas between Liberty National Golf Club and the New Jersey City University, with a tendency towards the northeast. The average elapsed time for the significantly detected crimes is $802.7$ seconds. Of the $156$ significantly detected crimes, $86\ (55.1\%)$ fall outside their corresponding envelopes even at small distance values, $r_{\LL} \leq 2000$, indicating strong local associations within their surroundings. To better understand the distance ranges for which these crimes are identified as having significant mark associations with their neighbouring crimes, we get all the ranges of $r_{\LL}$ for which the corresponding local mark correlation functions fall outside the envelope under the assumption of random labelling. These are shown in the right plot of Figure \ref{fig:jcr} where one can see that most significant ranges point to distances less than $r_{\LL} \leq 3000$.

\subsection{Pfynwald data}

The dataset comprises tree measurements collected annually over 14 years as part of a long-term irrigation experiment in \textit{Pfynwald}, located in the central region of the \textit{Pfyn-Finges} National Park in Switzerland \citep{pfynwald:2016}. Launched in $2003$, this study aimed to evaluate how increased water availability affects individual trees and the broader ecosystem within a naturally dry Scots pine (Pinus sylvestris L.) forest. 
The dataset, available under an Open Database License\footnote{\url{https://opendata.swiss}}, includes spatial coordinates, initial treatment or control group assignments, and various characteristics for $900$ trees. For this analysis, we primarily focus on the annual total crown defoliation (TCD) and the precise locations of individual trees. TCD is a commonly used metric in forest monitoring to measure needle or leaf loss relative to a local reference tree. Some trees did not have the mark values for the entire study period, i.e., $2003-2016$. The cleaned data is a point pattern with $741$ points and function-valued marks, showing the yearly TCD from $2003$ to $2016$. We used the convex hull of the data as the corresponding window. Previously,  \cite{Eckardt2023MultiFunctionMarks} studied the cross-mark association between the average TCD and local pair correlation function as function-valued marks by applying the cross-type mark correlation functions. 

\begin{figure*}[t]
    \centering
    \includegraphics[scale=0.1]{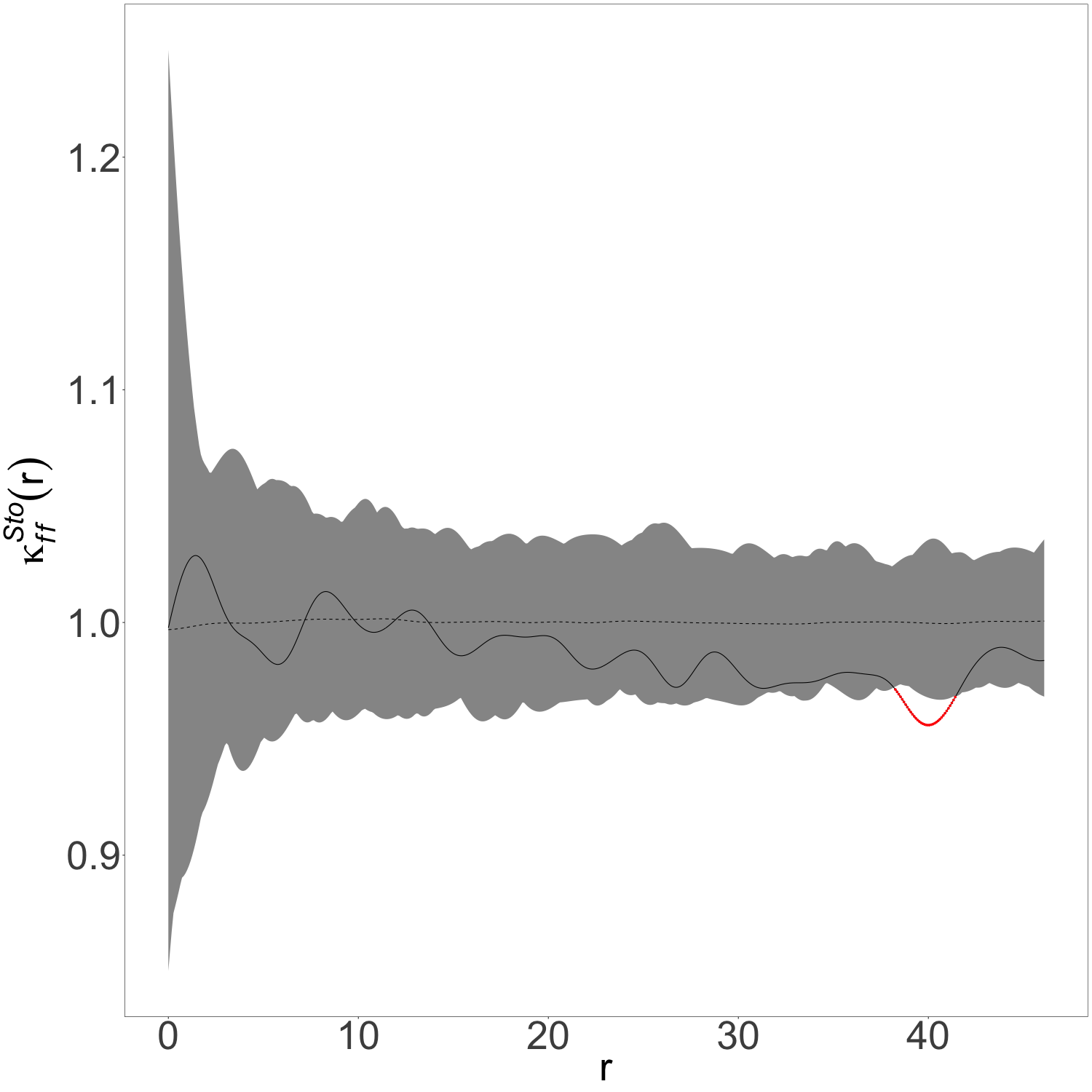}
    \includegraphics[scale=0.1]{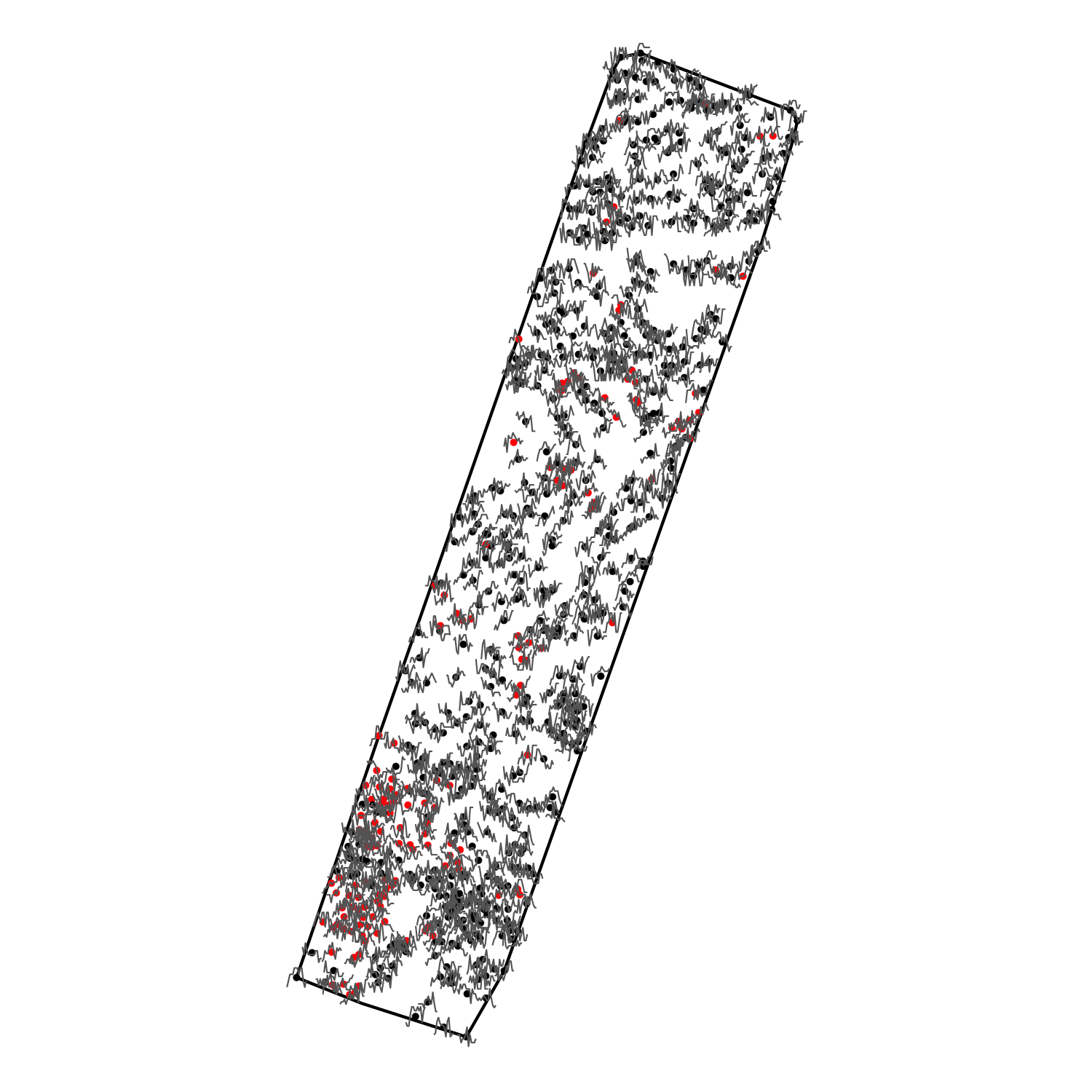}
    \includegraphics[scale=0.1]{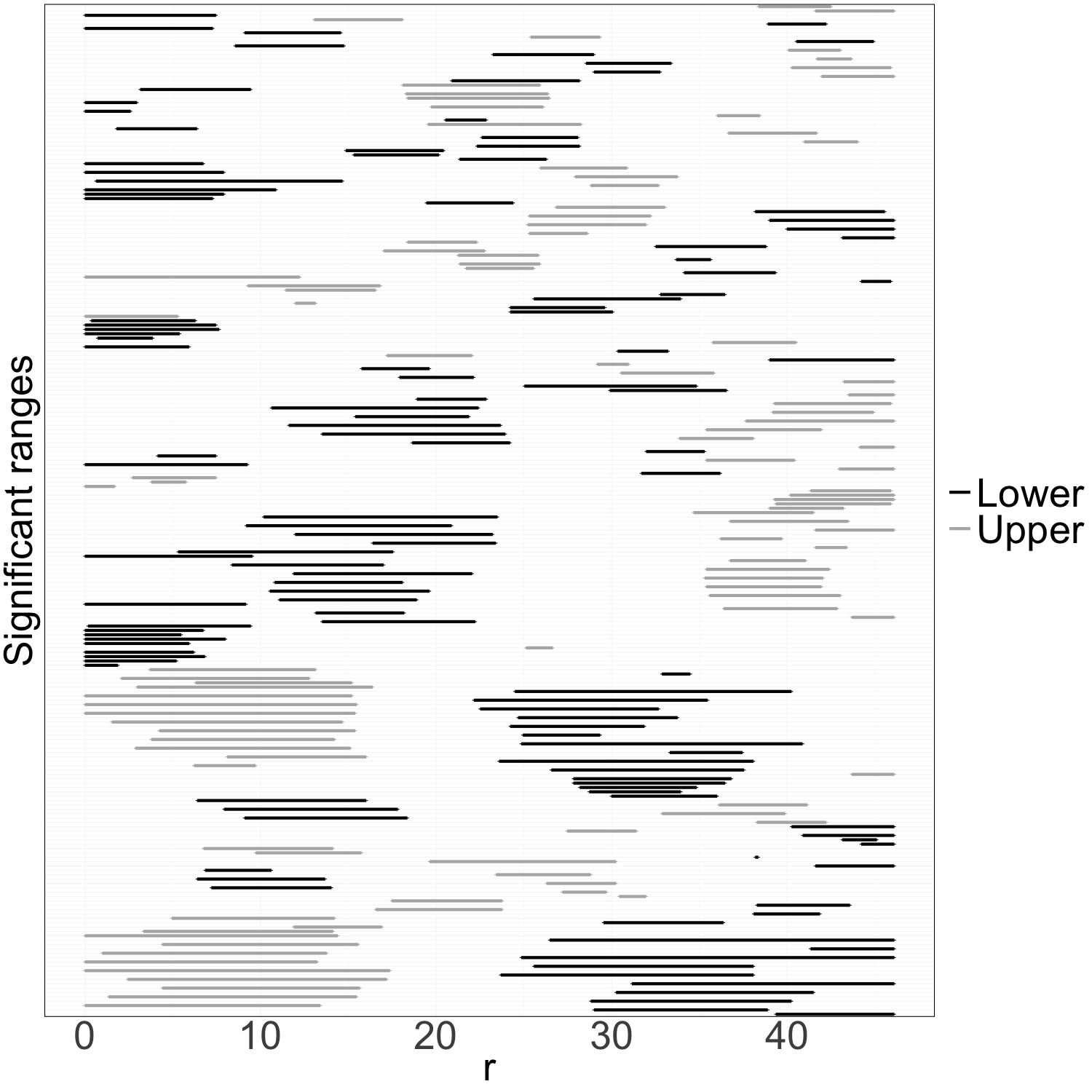}
    \caption{
    Results for Pfynwald data.
    Left: Global envelope test using the global mark correlation function $\kappa^{\mathrm{Sto}}_{ff}$.
    Middle: The point pattern of Pfynwald trees, with significant ones highlighted in red; the curves at each point show their function-valued mark. 
    Right: Ranges of distances $r$ for which the local mark correlation function $\kappa^{\mathrm{Sto}}_{f_if_j}$ falls outside the envelope for the trees with significant mark associations represented in red in the middle plot. Lower and Upper refers to whether $\kappa^{\mathrm{Sto}}_{f_if_j}$ is falling outside the lower or upper bound. 
    }
    \label{fig:Pfy}
\end{figure*}

The middle panel of Figure \ref{fig:Pfy} shows the point pattern of the trees and their corresponding function-valued marks as grey curves on top of points. Similar to our procedure for the previous two real datasets, we employ the global mark correlation function $\kappa^{\mathrm{Sto}}_{ff}$ jointly with global envelope tests based on $500$ permutations; the corresponding $p$-value is $0.014$. Looking at the obtained global envelopes for the $\kappa^{\mathrm{Sto}}_{ff}$, presented in the left panel of Figure \ref{fig:Pfy}, we can see that for small and moderate values of distance $r$, it fully stays inside the envelope, showing no deviations from random labelling. For larger values of $r \in (38.26, 41.41)$, it, however, goes outside the envelope from its lower bound.  This means that for pairs of trees with an interpoint distance $r \in (38.26, 41.41)$, the product of marks is, on average, smaller than the corresponding value under mark independence, which in turn means that for these pairs of trees, at least one of them has a very small mark. At the same time, we should remember that this may not be the true pattern of mark associations among trees, as domination can easily happen between distinct mark behaviours. 
This being said, the global mark correlation function $\kappa^{\mathrm{Sto}}_{ff}$ can not uncover more information about the distribution of TCD. Thereafter, we employ our proposed local mark correlation function $\kappa^{\mathrm{Sto}}_{f_i f_j}$, according to which we found that $20.5\%$ of the trees have significant local contributions at a significance level of $0.05$. The trees with significant contributions to the distributions of TCD are highlighted in red in the middle panel of Figure \ref{fig:Pfy}; interestingly, we can see that most of such trees are located at the bottom left of the study area. After checking the global envelope tests for all the significant trees, we noticed that for the significant ones in the bottom left, the local mark correlation function $\kappa^{\mathrm{Sto}}_{f_i f_j}$ stays outside of the envelope for small/moderate values of $r$ for most of them. For some other trees  $\kappa^{\mathrm{Sto}}_{f_i f_j}$ stays outside of the envelope for large values of $r$. These significant behaviours/findings were not uncovered when employing the global mark correlation function $\kappa^{\mathrm{Sto}}_{ff}$. To better see the ranges of interpoint distances $r$ for which the trees highlighted in red have been identified as having significant mark associations with neighbouring trees, we show such ranges of distances in the right panel of Figure \ref{fig:Pfy}. One can see that for the majority of significant ones, and $r \leq 20$, $\kappa^{\mathrm{Sto}}_{f_if_j}$ falls outside the envelope from the above (grey intervals); this was not detected at all by the global mark correlation function  $\kappa^{\mathrm{Sto}}_{ff}$. Interestingly, when $r \geq 35$, for most trees with significant contributions, their $\kappa^{\mathrm{Sto}}_{f_if_j}$ fall outside from the upper bound, but apparently the contributions of those trees for which $\kappa^{\mathrm{Sto}}_{f_if_j}$ fall outside from the lower bound has dominated the rest giving rise to the global mark correlation function   $\kappa^{\mathrm{Sto}}_{ff}$ falling outside the envelope from the lower bound; see the left panel of Figure \ref{fig:Pfy}.

\subsection{Urban mobility data}

In this application, we deal with a marked point pattern on a linear network for which marks are functions. More specifically, 
we study monthly bike-sharing data from Vancouver’s public bike-share program, \textit{Mobi by Shaw Go}, released under a public data license\footnote{\url{https://www.mobibikes.ca/en/system-data}}. Launched in $2006$, this system includes $250$ docking stations distributed city-wide, with stations positioned roughly every two to three blocks ($200$–$300$ meters apart). 
The dataset provides detailed information about the bike trips within a part of Vancouver, Canada, including the locations of departure and return stations, precise start and end timestamps, distance travelled (in meters), trip duration (in seconds), number and length of stops (interruptions), battery voltage at both departure and return points, and temperature at each docking station. 
Recently,  
\cite{Eckardt:Moradi:STAT} used average daily cycling distances as function-valued marks and compared the association among marks over a period of six months.
Here, we only focus on data from August $2022$; the dataset for this particular month includes $191$ points. In particular, the spatial points are the locations of bike stations, and function-valued marks are the average daily trip duration based on the departure station. 

\begin{figure*}[!h]
    \centering
    \includegraphics[scale=0.1]{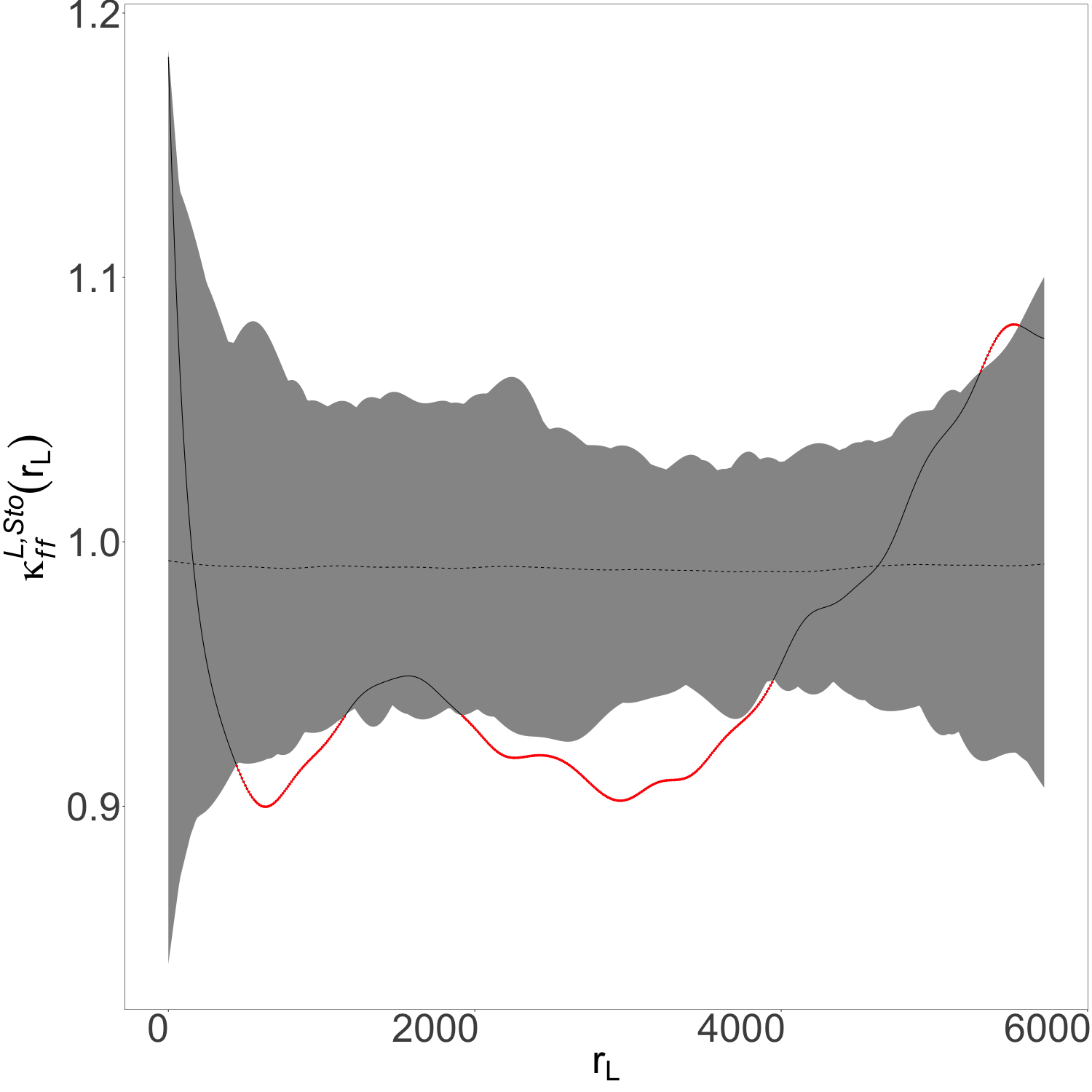}
    \includegraphics[scale=0.1]{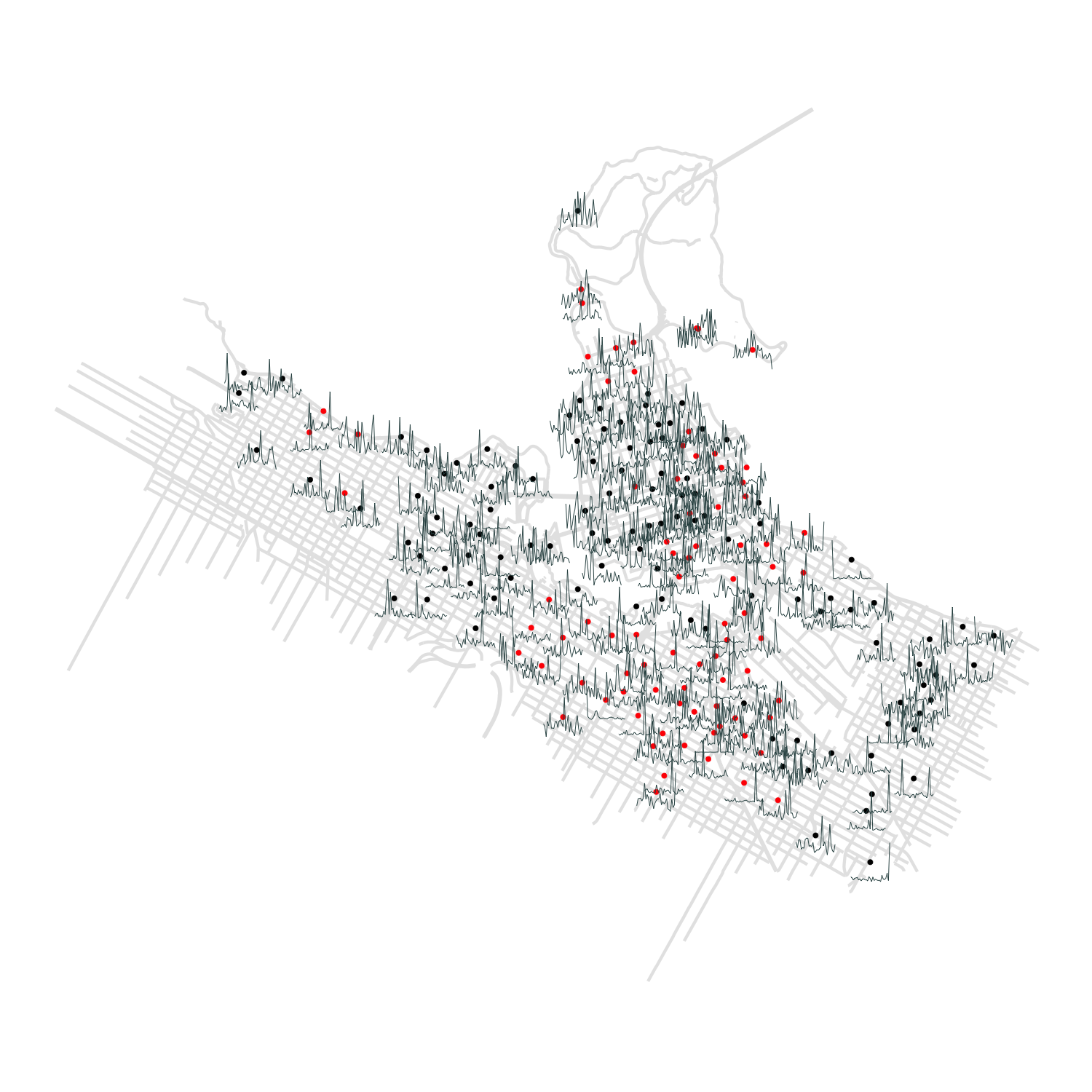}
    \includegraphics[scale=0.1]{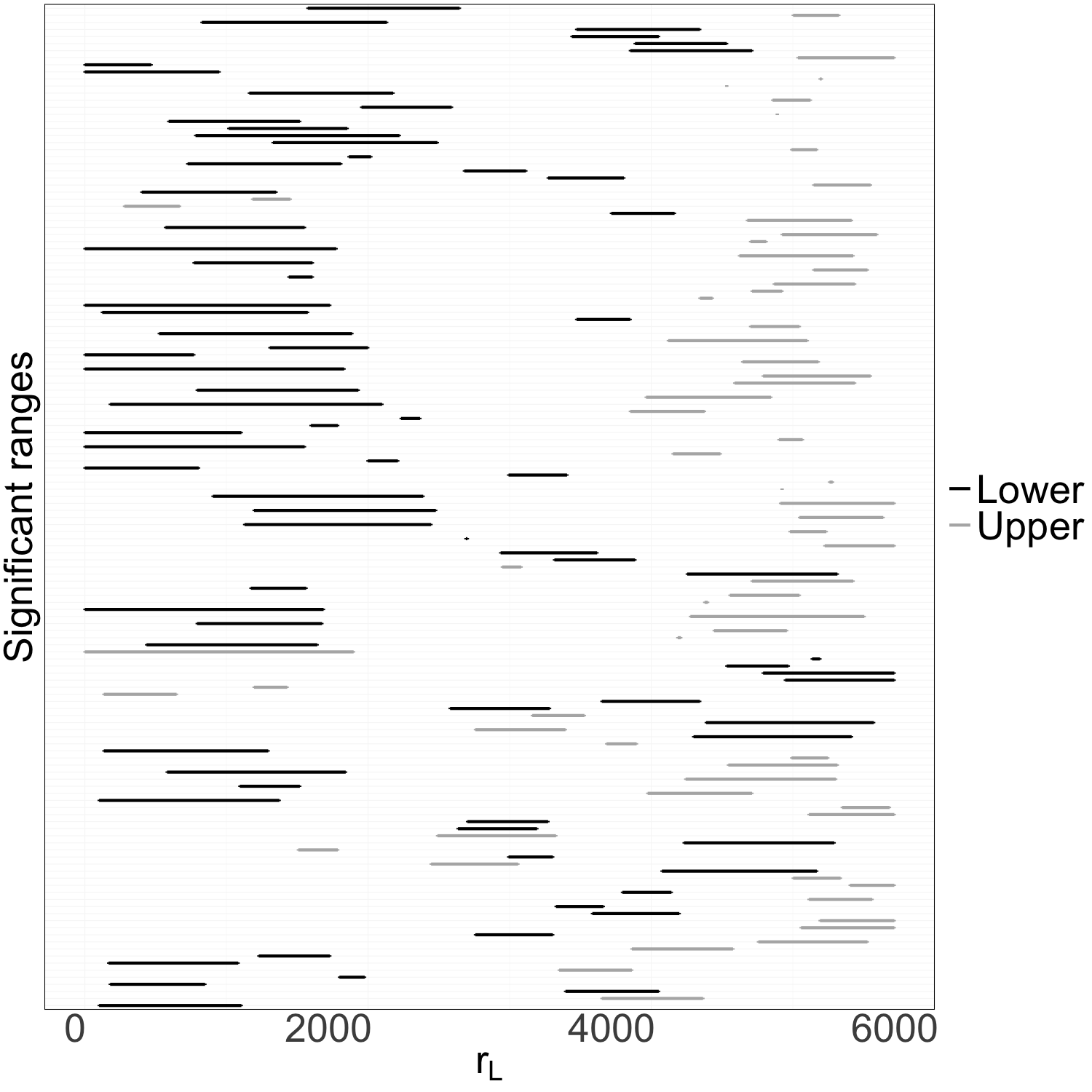}
    \caption{
    Results for urban bike sharing.
    Left: Global envelope test using the global mark correlation function $\kappa^{\mathrm{\LL, Sto}}_{ff}$.
    Middle: The point pattern of bike stations in Vancouver, with significant ones highlighted in red; the curves at each point show their function-valued mark. 
    Right: Ranges of distances $r_{\LL}$ for which the local mark correlation function $\kappa^{\LL, \mathrm{Sto}}_{f_if_j}$ falls outside the envelope for the bike stations with significant mark associations represented in red in the middle plot. Lower and Upper refers to whether $\kappa^{\LL, \mathrm{Sto}}_{f_if_j}$ is falling outside the lower or upper bound.
    }
    \label{fig:bikefig}
\end{figure*}

The middle panel of Figure \ref{fig:bikefig} represents the point pattern of the bike stations and their corresponding function-valued marks as grey curves on top of the points. Similar to our procedure for the previous three real datasets, we employ the global mark correlation function $\kappa^{\LL, \mathrm{Sto}}_{ff}$ jointly with global envelope tests based on $500$ permutations; the corresponding $p$-value is $0.002$. From the obtained global envelope, one can see that the $\kappa^{\LL, \mathrm{Sto}}_{ff}$ falls outside the envelope for small to moderate-size values of travelling distance $r_{\LL}$. In particular, for $r_{\LL} \in (446, 1150) \cup (1920, 3941)$ it falls outside from the lower bound, meaning that, on average, for bike stations with such interpoint distances, the product of marks is significantly less than that under mark independence, and for 
$r_{\LL} \in (5303, 5549)$ it falls outside the upper bound, meaning that for these interpoint distances, the product of marks is significantly higher than that under mark independence. 
Additionally, our findings indicate associations among the marks, with two distinct patterns emerging as the mark correlation function $\kappa^{\LL, \mathrm{Sto}}_{ff}$ falls outside the envelope's bounds depending on interpoint distances $r_{\LL}$. However, these patterns cannot be generalized across all bike stations, and it remains unclear which stations contribute to the observed associations. Furthermore, there is still a risk that the behaviour shown in Figure \ref{fig:bikefig} is biased in the sense that it may not show the true pattern of association as some function-valued marks might dominate each other, and some behaviours might get masked. 

Aiming to uncover the individual bike stations that have significant associations with other stations in their surrounding, we now employ our proposed LIMA functions. Similarly, we make use of the local mark correlation function $\kappa^{\LL, \mathrm{Sto}}_{f_i f_j}$. Considering the significance level of $0.05$, it is found that $42.4 \%$ of the bike stations have significant associations with their neighbouring stations at some interpoint distance $r_{\LL}$. These stations are highlighted in red in the middle plot of  Figure \ref{fig:bikefig}, from which we can see that the majority of bike stations with significant mark associations in terms of trip duration are located in the streets in the north of Queen Elizabeth Park, with some other bike stations, detected as having significant mark associations, located in downtown Vancouver. 

Next, we look into the details of each individual bike station that is detected as significant. The right plot of Figure \ref{fig:bikefig} shows all ranges of travel distance $r_{\LL}$ for which $\kappa^{\LL, \mathrm{Sto}}_{f_if_j}$ falls outside corresponding envelopes; these ranges are in agreement with the overall behaviour detected by the global mark correlation function $\kappa^{\LL, \mathrm{Sto}}_{ff}$ displayed in the right plot. However, the advantage of using LIMA functions is that they allow us to pinpoint which bike stations and within what travel distance ranges show significant associations regarding average daily trip durations.

\section{Discussion}\label{sec:diss}

When the spatial distribution of marks varies largely, which is especially the case when the observation window is large, it is not uncommon to have different mark behaviours that may overshadow one another, potentially leading to inaccurate conclusions based on global summary statistics and obscuring the complete association patterns among marks.
Here, we have developed the class of LIMA functions, integrating various local mark correlation functions into a unified framework.
More specifically, we have considered two cases of marked point processes: those on planar spaces and those on linear networks, with marks that are either real-valued or function-valued.
These LIMA functions enable us to extract the contributions of individual points to the overall spatial distributional behaviour of marks, leading to a complete understanding of the mark association/variation among marks. 


Through simulation studies across various scenarios, we observed that our proposed LIMA functions consistently outperform their global counterparts in detecting spatial patterns within marks. These findings highlight the enhanced sensitivity and precision of our LIMA functions in capturing localized mark structures, making them a superior tool for spatial analysis of complex marked point processes. We have observed that, based on global envelope tests \citep{myllymaki2017global}, our proposed LIMA functions exhibit a slightly lower type I error rate compared to their global counterparts while demonstrating greater power in detecting associations/variations among marks. Additionally, global mark correlation functions are limited by the tendency of distinct mark behaviours to dominate each other, obscuring the true pattern of mark structure. Even when global mark correlation functions identify mark associations, they fall short in pinpointing the specific regions or individual points contributing to these associations. In contrast, our LIMA functions effectively highlight the points that significantly contribute to observed mark structures and associations, offering a localized perspective and, consequently, a more detailed/complete understanding of mark associations compared to their global counterparts. Alongside its inherent contributions, our proposed framework can enhance the specification of spatial marked point process models by taking the local properties into account rather than completely relying on global ones. 

\section*{Acknowledgement}
 Matthias Eckardt has been supported by the German Research Foundation through Walter Benjamin grant 467634837.

\bibliographystyle{ecta}
\bibliography{LIMA}
\end{document}